\documentclass[12pt]{article}
\usepackage{dcolumn}
\usepackage{bm}
\usepackage{makeidx}
\usepackage{amssymb}
\usepackage{amsfonts}
\usepackage{amsmath}
\usepackage{graphicx}
\usepackage{setspace}
\usepackage{authblk}
\usepackage{units}
\usepackage{appendix}
\usepackage[left=2cm,top=2.3cm,right=2cm,bottom=2.1cm]{geometry}
\usepackage[hyperfootnotes=false]{hyperref}

\newcommand{\bee}{\begin{eqnarray}}
\newcommand{\eend}{\end{eqnarray}}

\newcommand{\bea}{\begin{eqnarray}}
\newcommand{\eea}{\end{eqnarray}}

\renewcommand\familydefault\rmdefault

\hypersetup{colorlinks=true,linkcolor=black,citecolor=black,urlcolor=black}
\begin{document}

\title{\textbf{Axiomatic quantum electrodynamics: from causality to
convexity of effective action }}
\author{Anatoly E. Shabad}
\affil { P.N. Lebedev Physics Institute, Moscow 117924, Russia}

\maketitle

\begin{abstract}
Theory of electromagnetic field, specified by an effective action
functional, is considered. The causality condition is imposed in the form of
a requirement that the group velocities of propagation of small and soft
disturbances over the background of an external constant field should not
exceed the speed of light in vacuum. It is shown that these conditions lead,
in particular, to a very definite conclusion about the geometry of the local
limit of the effective action. Namely, the surface, which is specified in
the local limit by the nonlinear Lagrangian, considered as a function of two
invariants of the field has positive Gaussian curvature.
\end{abstract}



\section{Introduction\label{Sec1}}

\onehalfspacing

In theories with scalar fields, stable states provide minimum to the action,
and excitations over such states propagate slower than light, as they
should. The action responsible for self-interaction is a concave (of
downward convexity) functional of the field. On the contrary, unstable
states pregnant with spontaneous symmetry breakdown provide maximum to
action, and excitations over them -- until the symmetry is broken -- are
causality violating superluminal agents -- tachyons. (This makes the
symmetry violation a categorical imperative).

The matter is more sophisticated when a vector field is concerned. Since the
very time of creation of Quantum Electrodynamics (QED) it is known \ that
electromagnetic fields interact with themselves, this interaction being
described by a (generally nonlocal) effective action, responsible for the
nonlinearity and serving as the generating functional of many-photon
vertices \cite{weinberg}. The origin of nonlinearity lies in the
light-by-light scattering peculiar to QED due to the quantum process of
virtual electron-positron pair production by a photon \cite{BLP}. By now the
existence of photon self-interaction has been confirmed experimentally in
peripheral nucleus-nucleus collisions \cite{atlas}.

In \cite{convexity} and \cite{Shabus2011}, Usov and the present author
considered a general electromagnetic field theory and excluded the
superluminal propagation of (small-amplitude and small-momentum) wave
packets over a constant magnetic field background (magnetized vacuum). When
joined with the requirement that residues of the photon propagator in the
mass-shell poles be nonnegative (referred to in those publications, for
brevity, as the unitarity principle), this resulted in the fact that the
local limit $\mathfrak{L}(\mathfrak{F},\mathfrak{G})$ of effective nonlinear
Lagrangian$,$ that gives rise to field equations of motion via the least
action principle, \textbf{\ }is a convex function of two invariants of
electromagnetic field \footnote{%
where the dual field tensor is defined as $\tilde{F}_{\rho \sigma }=\frac{1}{%
2}\epsilon _{\rho \sigma \lambda \kappa }F^{\lambda \kappa }$, with the
completely antisymmetric unit tensor defined in such a way that $\epsilon
_{1230}=1.$ Greek indices span the 4-dimensional Minkowski space taking the
values 1,2,3,0, the metric tensor is $\eta _{\rho \nu }=$ diag $(1,1,1,-1)$,
and $\square =\nabla ^{2}-\partial _{0}^{2}$.} $\mathfrak{F}=\frac{1}{4}%
F_{\rho \sigma }F^{\rho \sigma }$ and $\mathfrak{G}=\frac{1}{4}F^{\rho
\sigma }\tilde{F}_{\rho \sigma }$ if taken at zero\ value of one of them,$\
\mathfrak{G=}$ $0$:%
\begin{equation*}
\left. \text{ }\frac{\partial ^{2}\mathfrak{L}(\mathfrak{F},\mathfrak{G})}{%
\partial \mathfrak{F}^{2}}\right\vert _{\mathfrak{G=}0}\geqslant 0,\left.
\text{ }\frac{\partial ^{2}\mathfrak{L}(\mathfrak{F},\mathfrak{G})}{\partial
\mathfrak{G}^{2}}\right\vert _{\mathfrak{G=}0}\geqslant 0.
\end{equation*}%
This one and other consequences derived from the above requirements in Refs.
\cite{convexity} and \cite{Shabus2011} proved to be useful in providing
criteria for selection of appropriate models, not necessarily associated
with QED, when nonlinear electrodynamics is combined with the Einstein
gravity in search of new scenarios in cosmology and new black hole families
(see e.g. \cite{bronnikov} -- \cite{kkrugl} and references therein), also
when the Lagrangian of nonlinear electrodynamics is used for parametrization
of mutual scattering of photons produced in heavy ion collisions \cite%
{medeiros}, and in other instances.

To meet the demands of application, in the present paper we extend
consequences of causality to a more general case when the invariant $%
\mathfrak{G}$ is nonzero, $\mathfrak{G\neq }$ $0$. To be more precise, we
study propagation of small-amplitude electromagnetic waves against the
background of a constant electromagnetic field of the most general form,
i.e. the one where the both field invariants are different from zero, $%
\mathfrak{F}$ $\neq 0,$ $\mathfrak{G}$ $\neq 0.$ In the limit of vanishing
momentum components of these waves we find that the ban of superluminal
signals (combined also with the newly claimed "Maxwell Dominance Principle"
(MDP), saying that for not too strong nonlinearity\ the linear
electrodynamics of Faraday-Maxwell should dominate) results in that the
two-dimensional surface given by the function $Z=\mathfrak{L}(\mathfrak{F},%
\mathfrak{G})$ should have its Gaussian curvature, defined by the Hesse
determinant, nonnegative,
\begin{equation}
\mathfrak{L}_{\mathfrak{FF}}\mathfrak{L}_{\mathfrak{GG}}-\mathfrak{L}_{%
\mathfrak{FG}}^{2}\geqslant 0.  \label{Hesse}
\end{equation}%
(The subscripts $\mathfrak{F}$ and $\mathfrak{G}$ denote partial derivatives
with respect to the corresponding variables, both of them being now,
generally, nonzero). This surface is tangent to the coordinate plane $%
\mathfrak{\ }Z\mathfrak{\ }=0$ in the origin $\mathfrak{F}$ $=$ $\mathfrak{G=%
}$ $0,$ owing to the relations $\mathfrak{L}(0,0)=0,$ $\mathfrak{L}_{%
\mathfrak{F}}(0,0)=\mathfrak{L}_{\mathfrak{G}}(0,0)=0$ that are to be
imposed in concord with the requirement that the Maxwell-Faraday linear
electrodynamics should be reproduced in the asymptotic limit of small
fields. This is the Correspondence principle, not to be confused with MDP,
which may be thought of as an extension of the former. The above postulates
also yield other inequalities between derivatives of $\mathfrak{L}$ ,
namely, $\mathfrak{L}_{\mathfrak{FF}}\geqslant 0,$ $\mathfrak{L}_{\mathfrak{%
GG}}\geqslant 0$ again and also certain inequalities containing
fields explicitly. As for the positivity of the photon Green
function residues, it appears to be one of the consequences of MDP.

In the present paper, the same as in \cite{convexity}, \cite{Shabus2011}, we
are basing our causality argumentation on the\ notion of the group velocity
as the one that is responsible \cite{BornWolf} for the speed of a wave
packet, which may be used as an information-transferring signal and whose
speed, therefore, must not exceed the speed of light in the vacuum $c$\
(taken equal to unity throughout the paper) according to the basic postulate
of the theory of relativity \footnote{%
It is often pointed \ that the group velocity may exceed the speed
of light in the case of so-called abnormal dispersion. However, this
objection was lifted in \cite{ShabUs2010} by considering the real
part of the complex group velocity as a function of real momentum.}.
\ Certainly, there exist other approaches aimed to incorporate
causality on the general basis. The causal view that electric
induction is permitted to depend on electric field taken only in
past time moments gave rise to Kramers-Kronig dispersion relations
\cite{kr} -- \cite{Kramers} (these are also applicable to vacuum
with a background field, expressing nonlinearity of
electromagnetism). Whereas this idea has a nonlocal character and
does not concern directly the local Lagrange function
$\mathfrak{L}(\mathfrak{F},\mathfrak{G}),$ an approach of M. Novello
\emph{et al}. \cite{novello1} that goes back to Hadamard's
\cite{Hadamard} notion of electromagnetic field discontinuities
propagating in a given metric might well produce limitatiions on $\mathfrak{L%
}(\mathfrak{F},\mathfrak{G})$. Indeed, according to \cite{novello1}, \cite%
{Novello2} photons propagate along null geodesics of an effective
metric formed by $\mathfrak{L}(\mathfrak{F},\mathfrak{G})$, its
derivatives and field invariants (for later developments see
\cite{Amanda} and references therein)$.$ It is important to note
that different causal limitations, obtained following independent
procedures, if correct, may either overlap or complete each other.
In Appendix below we deal with an important example of such
situation.

It should be stressed that unlike the phase velocity \ $v_{ph}$ = $\frac{%
k_{0}}{|k|}$, which is, generally, not the speed of information transfer,
the group velocity is a 3-vector. Its components are defined as the
frequency differentiated over components of the wave 3-vector
\begin{equation}
v_{i}^{\mathrm{gr}}=\frac{\partial k_{0}}{\partial k_{i}},\qquad i=1,2,3
\label{2}\end{equation}%
and taken -- for each polarization mode -- on solution of the
corresponding dispersion equation. In other words,
$v_{i}^{\mathrm{gr}}$ is
a tangent vector to the surface defined by the dispersion equation $%
k_{0}=k_{0}(k_{i}). $ There is a vast discussion in the literature on
properties of the wave packet propagation, completed by experiments
\footnote{%
Nevertheless, an analysis of issues involving the group velocity in notions
of differential geometry is absent. Ours is not one, either.}. For our
present purposes, decisive is the fact established in \cite{Shabad2016}
(also discussed in \cite{Fresneda}) that under a Lorentz boost the group
velocity is transformed following the standard relativistic law $\oplus $ of
speed addition
\begin{equation}
v_{\parallel }^{\prime }{}^{\mathrm{gr}}=v_{\Vert }^{\mathrm{gr}}\oplus
\mathbf{V}\equiv \frac{V+v_{\Vert }^{\mathrm{gr}}}{1+Vv_{\Vert }^{\mathrm{gr}%
}},\text{ \ \ \ }\boldsymbol{v}_{\perp }^{\prime }{}^{\mathrm{gr}}=%
\boldsymbol{v}_{\perp }^{\mathrm{gr}}\oplus \mathbf{V}\equiv \frac{%
\boldsymbol{v}_{\perp }^{\mathrm{gr}}\left( 1-V^{2}\right) ^{1/2}}{%
1+Vv_{\Vert }^{\mathrm{gr}}},  \label{addition}
\end{equation}%
where the subscripts $\parallel $ and $\perp $ mark projections onto
directions, respectively, parallel and orthogonal to the speed $\mathbf{V}$
of the moving (primed) frame.\ Derivation of Eq. (\ref{addition}) is based
upon the fact that eigenvalues of the polarization tensor, that define
dispersion equations, are Lorentz scalars. The law (\ref{addition}) will
allow us to restrict the group velocity to below unity in a convenient
reference frame in understanding that it will remain, thereby, underluminal
in any other inertial frame. This convenient frame is the one where the
electric and magnetic parts of the constant background field are mutually
parallel, when $\mathfrak{G>}$ $0$, or antiparallel, when $\mathfrak{G<}$ $0$%
.

In Section 2\ we discuss a sort of axiomatic approach to the Abelian gauge
theory of a vector (electromagnetic) field basing upon an \emph{a priory}
knowledge of the effective action as of a (nonlocal) functional of the
electromagnetic field tensor, without referring to a way it is to be
calculated within one or another definite dynamical theory, say QED. Knowing
this object implies the knowledge of all nonlinear (integro-differential)
equations of motion, since the nonlocal effective action generates via
variational differentiations all many-photon vertices (\cite{weinberg}),
that serve as kernels to these equations. Moreover, being the Legendre
transform of the generating functional of Green functions, it can be used
for reproducing also the latter \cite{Peskin}. This approach implies an
exploration of obligatory properties prescribed to the effective action and
to many-photon vertices by fundamental principles alone.

As an example, we present the results concerning the two-photon vertex --
second-rank polarization tensor in a background field -- governing the
process of light propagation linearized near that field. Taking the field
with its strength constant in time and space -- both invariants $\mathfrak{F}
$ and $\mathfrak{G}$ being nonvanishing -- for the background, we can make
some statements concerning the photon polarization eigenmodes and their
dispersion laws, which are grounded only on relativistic, gauge and
translation invariance and independent of any dynamical model and of any
approximation within it (\cite{batalin}), (\cite{Trudy}). Certain
predictions can be formulated about dispositions of dispersion curves \cite%
{ShabUs2010} $k_{0}=k_{0}(k_{i})$ for any $\mathbf{k}$, that follow also\
from the other fundamental principle -- the causality. The frequency
(energy) of a photon in each mode $k_{0}$ is, generally, a sophisticated
function of its momentum (wave vector) $\boldsymbol{k}$\ due to nonlocality
of the effective action functional. For this reason we are at the present
stage of work unable to efficiently handle the group velocity approach in
the general case to obtain causality restrictions on the action in terms of,
expectably, its variational derivatives. However, the function $%
k_{0}=k_{0}(k_{i})$ can be found explicitly in the limit\footnote{%
The point $k_{0}=$ $\boldsymbol{k}=0$ belongs to the photon dispersion
curve, since the massless particle should have vanishing energy, $k_{0}=$ $%
0, $ at rest $\boldsymbol{k}=0$. Massive branches with the endpoint $%
k_{0}\neq 0,$ $\boldsymbol{k}=0$ may also exist, but we do not exploit them.}
$k_{0}\rightarrow 0,$ $\boldsymbol{k}\rightarrow 0,$ served by the local
limit of the effective action, where it is -- like, for instance, the
Born-Infeld or Euler-Heisenberg actions -- a functional of fields and not of
their space- and time-derivatives.

In Section \ref{Sec3} we focus on constraints imposed on the local limit of
the effective action by the requirement that the steady long-wave
excitations above the most general constant background be not faster than
light. It ensues that in the special frame, where the magnetic and electric
background fields \ are parallel or antiparallel the dispersion curves for
both the ordinary and extraordinary waves in the limit $k_{0}\rightarrow 0,$
$\boldsymbol{k}\rightarrow 0$\ are straight lines in the plane ($%
k_{0}^{2}-k_{\parallel }^{2},k_{\perp }^{2}),$ where $k_{\parallel
},k_{\perp }$ are photon momentum components across and along the common
direction of the background fields, their slope angles ranging inbetween%
\footnote{%
These straight lines are the asymptotes near $k_{0}\rightarrow 0,$ $%
\boldsymbol{k}\rightarrow 0$ of the two true dispersion curves $%
k_{0}^{2}-k_{\parallel }^{2}=f(\boldsymbol{k}_{\perp }^{2}),$ created by
nonlocal action.} $\ 0$ and $\pi /2.$ In other words both lines point in the
space-like directions $k^{2}=\boldsymbol{k}^{2}-k_{0}^{2}\geqslant 0.$ This
configuration of dispersion curves traces back to a number of relations
between second partial derivatives of $\mathfrak{L}(\mathfrak{F},\mathfrak{G}%
),$ the geometric property (\ref{Hesse}) among them, which have been the
goal of the present paper.

In Conclusion, details of relationships between different principles and
their consequences are thoroughly traced.

In Appendix we revisit the degenerate case of magnetized vacuum $\mathfrak{G}%
=0$ to compare our procedure with what can be deduced from applying the
so-called energy conditions \cite{ellis}. We find that the Weak Energy
Condition is equivalent to the positivity of residues principle, while the
Dominant Energy Condition is apt to reproduce all the consequences of
causality 
but one. The missing relation is that of convexity, $\left. \text{ }\frac{%
\partial ^{2}\mathfrak{L}(\mathfrak{F},\mathfrak{G})}{\partial \mathfrak{F}%
^{2}}\right\vert _{\mathfrak{G=}0}\geqslant 0.$ To reproduce it, energy
conditions should have been completed by Maxwell dominance principle, i.e.
by restriction imposed on the strength of nonlinearity.

\section{Nonlocal action functional, and n-rank polarization tensors
generated by it}

\onehalfspacing

We define a theory of self-interacting Abelian vector gauge
(electromagnetic) field $A_{\alpha }(z)$\ as given by the nonlocal effective
action functional%
\begin{equation}
S=\int L(x)d^{4}x,\text{ \ }L(x)=-\mathfrak{F(}x\mathfrak{)+L(}x),  \label{S}
\end{equation}%
where the nonlinear part $\mathfrak{L(}x)$ of the full Lagrangian $L(z)$ is
a function of the electromagnetic field strength tensor $F_{\alpha \beta
}(x)=\partial ^{\alpha }A_{\beta }(x)-\partial ^{\beta }A_{\beta }(x)$ and
of all its space- and time-derivatives, $F_{\alpha \beta }^{\left( m\right)
}\left( z\right) ,$of $m$-th order, $m=0,1,2...,$ taken in Lorentz-invariant
combinations. The dependence of $S$ \ exclusively on the field strengths --
and not on potentials -- guarantees (in the present Abelian context) gauge
invariance of the theory.

The nonlinear part $\Gamma =\int \mathfrak{L}(x)d^{4}x$ of the action (\ref%
{S}) is a generating functional of all $n$- photon vertices -- polarization
tensors of $n$-th rank -- taken against the background of a classical
external field $\mathcal{A}^{\text{ext}}$:%
\begin{equation}
\Pi _{\mu _{1}\mu _{2}...\mu _{n}}^{(n)}(x_{1},x_{2},...x_{n})=\left. \frac{%
\delta ^{n}\Gamma }{\delta A_{\mu _{1}}(x_{1})\delta A_{\mu
_{2}}(x_{2})...\delta A_{\mu _{n}}(x_{n})}\right\vert _{A=\mathcal{A}^{\text{%
ext}}},\text{ \ \ }n=2,3,4...  \label{Pn}
\end{equation}%
so that equations of motion -- the full set of source-less nonlinear Maxwell
equations above the external field -- provided by the least action principle%
\begin{equation}
\frac{\delta S}{\delta A_{\mu }\left( x\right) }=0,  \label{least action}
\end{equation}%
takes the form of the Taylor expansion%
\begin{eqnarray}
\text{\ }\frac{\delta S}{\delta A_{\mu }\left( x\right) } &=&\left. \frac{%
\delta S}{\delta A_{\mu }\left( x\right) }\right\vert _{A=\mathcal{A}^{\text{%
ext}}}+\left[ \eta _{\mu \nu }\square -\partial ^{\mu }\partial ^{\nu }%
\right] a^{\nu }(x)+\int \left. \Pi _{\mu \nu }^{(2)}(x,x^{\prime
})\right\vert _{A=\mathcal{A}^{\text{ext}}}a^{\nu }(x^{\prime
})d^{4}x^{\prime }  \label{taylor} \\
&&+\frac{1}{2}\int \left. \Pi _{\mu \nu \lambda }^{(3)}(x,x^{\prime
},x^{\prime \prime })\right\vert _{A=\mathcal{A}^{\text{ext}}}a^{\nu
}(x^{\prime })a^{\lambda }(x^{\prime \prime })d^{4}x^{\prime }d^{4}x^{\prime
\prime }+...=0  \notag \\
&&\text{\ }  \notag
\end{eqnarray}%
in powers of the deviation $a_{\mu }(x)$\ of the electromagnetic field from
the background
\begin{equation*}
A_{\mu }(x)=\mathcal{A}_{\mu }^{\text{ext}}+a_{\mu }(x),\text{ \ \ \ \ }%
F_{\alpha \beta }^{\text{ext}}=\text{ }\mathcal{\partial }^{\alpha }\mathcal{%
A}_{\beta }^{\text{ext}}\mathcal{-\partial }^{\beta }\mathcal{A}_{\alpha }^{%
\text{ext}}.
\end{equation*}%
The free Maxwell term $\left[ \eta _{\mu \nu }\square -\partial ^{\mu
}\partial ^{\nu }\right] a^{\nu }(x)$ in (\ref{taylor}) comes from the
variation of the linear part $\ S_{0}=-\int \mathfrak{F}(x)d^{4}x$ \ of the
action.

In what follows we shall deal with background fields of constant intensity, $%
\frac{\partial F_{\alpha \beta }^{\text{ext}}}{\partial x_{\mu }}=0.$

\bigskip The fact that $L(x)$ depends on the 4-space-time coordinate $x_{\mu
}$ only implicitly, \emph{i.e.} through the fields, guarantees in this case
the fulfillment of the physical requirement of translation invariance of
photon vertices against constant background (effective homogeneity of the
space-time for photons), reflected in that the tensors (\ref{Pn}) depend on
differences $(z_{i}-z_{i+1}),$ $i=1,2...n-1$\ of their arguments and,
correspondingly, their Fourier-transforms contain delta-functions of sums of
momenta of all photons in a vertex showing their energy-momentum
conservation.\ The formal proof of this statement may be found in the first
section of \cite{Trudy}.

The same fact also provides the continuity relations with respect to every
argument and every index (the transversality in the momentum space) for \
polarization tensors of every rank \ \ \
\begin{eqnarray}
\frac{\partial }{\partial x_{\mu _{i}}^{i}}\Pi _{\mu _{1}...\mu _{i}...\mu
_{n}}^{(n)}(x_{1},...x_{i},...x_{n}) &=&\frac{\partial }{\partial x_{\mu
_{i}}^{i}}\left. \frac{\delta ^{n}\Gamma }{\delta A_{\mu
_{1}}(z_{1})...\delta A_{\mu _{i}}(x_{i})...\delta A_{\mu _{n}}(z_{n})}%
\right\vert _{A=\mathcal{A}^{\text{ext}}}  \notag \\
&=&\frac{\partial }{\partial x_{\mu _{i}}^{i}}\left. \frac{\partial }{%
\partial x_{\tau }^{i}}\frac{\delta ^{n}\Gamma }{\delta A_{\mu
_{1}}(z_{1})...\delta F_{\tau \mu _{i}}(x_{i})...\delta A_{\mu _{n}}(z_{n})}%
\right\vert _{A=\mathcal{A}^{\text{ext}}}=0,  \label{transversality}
\end{eqnarray}%
fulfilled due to anti-symmetricity of the variational derivative $\frac{%
\delta \Gamma }{\delta F_{\tau \mu _{i}}(x_{i})}$ under transmutation of
indices $\tau \leftrightarrow \mu _{i}$. The transversality (\ref%
{transversality}) guarantees invariance of every term in\ the expansion (\ref%
{taylor})\ under the gauge (gradient) transformation of the field $a^{\nu
}\rightarrow a^{\nu }+\partial _{\nu }\varphi $ with $\varphi (x)$ being an
arbitrary scalar function decreasing when $x_{\mu }\rightarrow \infty $ $.$
Note, however, that the constancy of the background is not, as a matter of
fact, necessary for the transversality.

There is yet another consequence of the ban of explicit inclusion of the
coordinate into the action. This is that the constant field does not require
any current. In other words, the first term in (\ref{taylor}) disappears:%
\begin{eqnarray}
\left. \frac{\delta S}{\delta A_{\beta }\left( x\right) }\right\vert _{A=%
\mathcal{A}_{\text{ext}}} &=&\frac{\partial }{\partial x_{\alpha }}\left.
\frac{\delta S}{\delta F_{\alpha \beta }\left( x\right) }\right\vert _{A=%
\mathcal{A}_{\text{ext}}}=0.  \label{zero-current} \\
&&,  \notag
\end{eqnarray}%
This holds, since $\left. \frac{\delta S}{\delta F_{\alpha \beta }\left(
x\right) }\right\vert _{A=\mathcal{A}_{\text{ext}}}$ does not depend on $x$
once $F_{\alpha \beta }^{\text{ext}}=const.$ Indeed, $S$ contains $F_{\alpha
\beta }(x)$ and all its derivatives, $F_{\alpha \beta }^{\left( m\right)
}\left( z\right) .$ After the variational derivative is calculated one
should set $F_{\alpha \beta }(x)=const,$ and $F_{\alpha \beta }^{\left(
m\right) }\left( z\right) =0$ for $m>0.$ So, there is no way for dependence
of $\left. \frac{\delta S}{\delta F_{\alpha \beta }\left( x\right) }%
\right\vert _{A=\mathcal{A}_{\text{ext}}}$ on $x,$ remembering also that no
explicit dependence of $S$ on $x$ has been admitted.

Now, with the account of (\ref{zero-current}) the nonlinear field equations (%
\ref{taylor}) for the vector-potential deviation from the constant-field
background take the form%
\begin{eqnarray}
&&\text{\ }\left[ \eta _{\mu \nu }\square -\partial ^{\mu }\partial ^{\nu }%
\right] a^{\nu }(x)+\int \left. \Pi _{\mu \nu }^{(2)}(x-x^{\prime
})\right\vert _{A=\mathcal{A}^{\text{ext}}}a^{\nu }(x^{\prime
})d^{4}x^{\prime }  \notag \\
&&+\frac{1}{2}\int \left. \Pi _{\mu \nu \lambda }^{(3)}(x-x^{\prime
},x^{\prime }-x^{\prime \prime })\right\vert _{A=\mathcal{A}^{\text{ext}%
}}a^{\nu }(x^{\prime })a^{\lambda }(x^{\prime \prime })d^{4}x^{\prime
}d^{4}x^{\prime \prime }+...=0  \label{Max}
\end{eqnarray}

\bigskip Here the second-rank polarization tensor $\left. \Pi _{\mu \nu
}^{(2)}(x-x^{\prime })\right\vert _{A=\mathcal{A}^{\text{ext}}}$\ governs
the photon propagation in the equivalent medium formed by the background
constant \ field, while the third-rank polarization tensor $\left. \Pi _{\mu
\nu \lambda }^{(3)}(x-x^{\prime },x^{\prime }-x^{\prime \prime })\right\vert
_{A=\mathcal{A}^{\text{ext}}}$\ is responsible for nonlinear processes of
photon splitting into two and two-photon merging into one in that medium.
Higher-rank tensors describe multiple mutual photon conversions, for
instance, photon-by-photon scattering$.$ In the rest part of the paper we
deal only with the linearized field equations, which corresponds to small
deviations $a^{\nu }(x)$\ propagating above the background. This means that
we consider equation (\ref{Max}) without the cubic and higher-powers terms%
\begin{equation}
\text{\ }\left[ \eta _{\mu \nu }\square -\partial ^{\mu }\partial ^{\nu }%
\right] a^{\nu }(x)+\int \left. \Pi _{\mu \nu }^{(2)}(x-x^{\prime
})\right\vert _{A=\mathcal{A}^{\text{ext}}}a^{\nu }(x^{\prime
})d^{4}x^{\prime }=0.  \label{lineq}
\end{equation}%
In momentum representation $\Pi _{\mu \tau }(k,p)=\delta ^{4}(k-p)\Pi _{\mu
\tau }(k)=\int d^{4}xd^{4}x^{\prime }$ $\left. \Pi _{\mu \nu
}^{(2)}(x-x^{\prime })\right\vert _{A=\mathcal{A}^{\text{ext}}}\exp
\{ik^{\alpha }x_{\alpha }-ip^{\alpha }x_{\alpha }^{\prime }\}$ this equation
is algebraic:%
\begin{equation}
\left[ \eta _{\mu \nu }k^{2}-k_{\mu }k_{\nu }-\Pi _{\mu \nu }(k)\right]
a^{\nu }(k)=0  \label{algeb}
\end{equation}

\subsection{Second-rank polarization tensor}

The second-rank polarization tensor in a constant external field with its
two invariants \ $\mathfrak{F}^{\text{ext}}$ and $\mathfrak{G}^{\text{ext}}$
different from zero was studied on approximation-independent basis\footnote{%
This quantity also was calculated in these references within Quantum
Electrodynamics as a loop of electron and positron propagators for which
solutions to the Dirac equation in arbitrary constant field were
substituted, the Furry picture. We do not appeal to these calculations in
the present context, and exploit only model-independent conclusions of those
works.} in \cite{batalin}, \cite{Trudy} in QED.

It follows from the representation (\ref{Pn}) \ with $n=2$ and
translation invariance, that $\Pi _{\mu \tau }(k,A)=\Pi _{\tau \mu
}(-k,A).$ When coupled with the relation $\Pi _{\mu \tau }(k,A)=\Pi
_{\mu \tau }(-k,-A),$ which expresses invariance under full
space-time reflection (PT-), this produces $\Pi _{\mu \tau
}(k,A)=\Pi _{\tau \mu }(k,-A).$When now the Furry theorem
(C-invariance) $\Pi _{\mu \tau }(k,A)=\Pi _{\mu \tau }(k,-A)$ is
applied we come to the symmetricity of the polarization tensor $\Pi
_{\mu \tau }(k,A)=\Pi _{\tau \mu }(k,A)$ as the CPT-invariance
property. (So $\Pi _{\mu \tau }$ remains symmetric if the spacial
parity is violated together with T-parity, with C- and
PT-invariances kept valid.)

As any symmetric transversal tensor, $\Pi _{\mu \tau }$ has the following
diagonal representation

\begin{equation}
\Pi _{\mu \tau }(k,p)=\delta ^{4}(k-p)\Pi _{\mu \tau }(k)\,,\ \ \Pi _{\mu
\tau }(k)=\sum_{c=1}^{3}\varkappa _{c}(k)\frac{\flat _{\mu }^{(c)}\flat
_{\tau }^{(c)}}{(\flat ^{(c)})^{2}}  \label{diag}
\end{equation}%
in terms of its three mutually orthogonal $\flat _{\mu }^{(c)}\flat ^{(a)\mu
}=\delta _{ac}$, transversal $\flat _{\mu }^{(c)}k^{\mu }=0,$ real
eigenvectors $\flat _{\mu }^{(c)},$ and three scalar eigenvalues $\varkappa
_{c}(k)$ determined by solution of the eigenvalue problem%
\begin{equation}
\Pi _{\mu \tau }(k)\flat ^{(c)\tau }=\varkappa _{c}(k)\flat _{\mu }^{(c)},%
\text{ }c=1,2,3.  \label{eigenproblem}
\end{equation}%
The fourth eigenvector $\flat _{\mu }^{(4)}=k_{\mu }$ corresponds to
vanishing eigenvalue $\varkappa _{4}(k)=0$ due to the transversality $\Pi
_{\mu \tau }(k)k^{\tau }=0,$ and it does not participate in decomposition (%
\ref{diag}). Every product $a^{\nu }(k)=R_{c}(k)\flat _{\tau }^{(c)},$ where
$R_{c}(k)$ are arbitrary functions, is a solution to equation (\ref{algeb})
under the condition that the following dispersion equation-%
\begin{equation}
k^{2}-\varkappa _{c}(k)=0  \label{disp}
\end{equation}%
is satisfied. This implies that $\flat _{\tau }^{(c)}$ serves as the
amplitude of a vector potential of an eigen-mode number $c$ , governing its
polarization properties.

It was established \cite{batalin}\ that in momentum space the $4\times 4$
polarization tensor is a linear superposition
\begin{equation}
\Pi _{\mu \tau }(k)=\sum_{i=1}^{4}\Psi _{\mu \tau }^{\left( i\right) }\Theta
_{i}  \label{lincomb}
\end{equation}%
of four linearly independent transversal matrices (tensors) $\Psi _{\mu \tau
}^{\left( i\right) }$ formed by the tensor $F_{\alpha \beta }^{\text{ext}},$
the unit tensor $\eta _{\mu \nu }$, and the 4-momentum vector $k_{\mu }$ --
taken with four scalar coefficients $\Theta _{i}.$ Once the number of
coefficients in (\ref{lincomb}) is four, while the number of terms in (\ref%
{diag}) is three, it is clear that the eigenvalues $\varkappa _{c}(k)$ are
expressed via the coefficients $\Theta _{c}$ in an irrational way. An
exclusion is provided by the special, degenerate, case of
magnetic(electric)-like external field, $\mathfrak{G}_{\text{ext}}=0,$ $%
\mathfrak{F}$ $\gtrless 0$ where one of the four matrices becomes a linear
combination of others, and therefore there remain only three independent
matrices $\Psi _{\mu \tau }^{\left( i\right) }$. For this reason the
eigenvectors are found in a closed form as simple combinations of external
field tensor and photon momenta. We refer to this happening as
kinematization of the basis\footnote{%
This occurs in spacially-even case. See comments below Eq. (\ref{matrices}).}
when $\mathfrak{G}_{\text{ext}}=0$.\ Nevertheless, one of eigenvectors is
known independently of the coefficients $\Theta _{c}$ in the general case $%
\mathfrak{G}_{\text{ext}}\neq 0,$ too:
\begin{equation}
\flat _{\mu }^{(1)}=(F^{2}k)_{\mu }k^{2}-k_{\mu }(kF^{2}k)\,,\ \text{\ }%
\flat _{\mu }^{(1)}\flat ^{(1)\mu }=k^{2}\left[ k^{2}\left( k^{2}\mathfrak{G}%
^{2}-2\mathfrak{F}(kF^{2}k)\right) -(kF^{2}k)^{2}\right] ,  \label{b1}
\end{equation}%
while the dispersion equation (\ref{disp}) for $c=1$ is $(1-\mathfrak{L}_{%
\mathfrak{F}})k^{2}=0.$ Hence, as seen from (\ref{b1}), mode 1 becomes a
pure gauge on its mass shell $k^{2}=0$, its vector potential being a
gradient carrying no electromagnetic field. Therefore, we are left with two
nontrivial polarization degrees of freedom $c=2,3$, subject to
birefringence. We proceed with studying their asymptotic spectra for small $%
k_{\mu }\rightarrow 0.$ This asymptote is served for by the local limit of
the action. The point is{\Huge \ }that in this limit the behavior of $\Pi
_{\mu \tau }(k)$ as a function of momenta degenerates to quadratic one.
(Generally, \ $\Pi _{\mu \tau }^{(n)}$ behaves as $k^{n}$ when $k\rightarrow
0).$

\section{Causal restrictions on the local limit of action functional\label%
{Sec3}}

\onehalfspacing

\subsection{\protect\fbox{}\protect\bigskip Local limit of second-rank
polarization tensor}

In the local limit the action $S$\ does not contain space- and
time-derivatives of the field tensor $F_{\mu \nu }.$ Once there are only two
relativistic invariants free of derivatives%
\begin{equation}
\mathfrak{F}=\frac{1}{4}F_{\rho \sigma }F^{\rho \sigma }=\frac{1}{2}\left(
\mathbf{B}^{2}-\mathbf{E}^{2}\right) \text{ \ and \ \ }\mathfrak{G}=\frac{1}{%
4}F^{\rho \sigma }\tilde{F}_{\rho \sigma }=\left( \mathbf{BE}\right)
\label{invariants}
\end{equation}%
\textbf{\ }this implies that $L$ may depend only on them. So we shall take
\begin{equation}
\mathfrak{L}(x)=\mathfrak{L}(\mathfrak{F(}x\mathfrak{)},\mathfrak{G(}x%
\mathfrak{)})  \label{call.L}
\end{equation}%
in \ref{S}). The positive sign in the rightmost equality in (\ref{invariants}%
) is tied to the choice $\epsilon _{1230}=1$ of the clue component of the
unit anti-symmetric Levi-Chivita tensor in the definition of the dual field $%
\tilde{F}_{\rho \sigma }=$\ $\frac{1}{2}\epsilon _{\rho \sigma \lambda
\kappa }F^{\lambda \kappa }.$\

Eq.(\ref{call.L}) indicates that in the local limit the theory possesses the
dual invariance in the sense that the both values $\mathfrak{F(}x\mathfrak{)}
$ and $\mathfrak{G(}x\mathfrak{),}$ and hence $\mathfrak{L}(x),$ are
invariant under the transformations $\mathbf{E\rightleftarrows }$ $i\mathbf{%
B.}$ We shall employ this property.

Parity, if imposed, would require that $\mathfrak{L}$ be an even function of
the pseudoscalar $\mathfrak{G.}$ We do not keep to this assumption, though$%
\mathfrak{.}$

Then calculation of $\ \left. \Pi _{\mu \nu }^{(2)}(x-x^{\prime
})\right\vert _{A=\mathcal{A}^{\text{ext}}}$ following (\ref{Pn})
gives\bigskip\ \cite{Costa et al.}%
\begin{align}
\left. \Pi _{\mu \nu }^{(2)}(x-x^{\prime })\right\vert _{A=\mathcal{A}^{%
\text{ext}}}=& \frac{\delta ^{2}\Gamma }{\delta A^{\mu }(x)\delta A^{\tau
}(y)}=\int \mathrm{d}^{4}z\left\{ \frac{\partial \mathfrak{L}(\mathfrak{F}%
(z),\mathfrak{G}(z))}{\partial \mathfrak{F}(z)}\left( \eta _{\mu \tau }\eta
_{\alpha \beta }-\eta _{\mu \beta }\eta _{\alpha \tau }\right) \right. +
\notag \\
& +\frac{\partial \mathfrak{L}(\mathfrak{F}(z),\mathfrak{G}(z))}{\partial
\mathfrak{G}(z)}\epsilon _{\alpha \mu \beta \tau }+\text{ \ \ \ \ \ \ \ \ }
\notag \\
& +\frac{\partial ^{2}\mathfrak{L}(\mathfrak{F}(z),\mathfrak{G}(z))}{%
\partial (\mathfrak{F}(z))^{2}}F_{\alpha \mu }(z)F_{\beta \tau }(z)+\frac{%
\partial ^{2}\mathfrak{L}(\mathfrak{F}(z),\mathfrak{G}(z))}{\partial (%
\mathfrak{G}(z))^{2}}\tilde{F}_{\alpha \mu }(z)\tilde{F}_{\beta \tau }(z)+
\notag \\
& +\left. \frac{\partial ^{2}\mathfrak{L}(\mathfrak{F}(z),\mathfrak{G}(z))}{%
\partial \mathfrak{F}(z)\partial \mathfrak{G}(z)}\left[ {F}_{\alpha \mu }(z)%
\tilde{F}_{\beta \tau }(z)+\tilde{F}_{\alpha \mu }(z){F}_{\beta \tau }(z)%
\right] \right\} \left( \frac{\partial }{\partial z_{\alpha }}\delta
^{4}(x-z)\right) \left( \frac{\partial }{\partial z_{\beta }}\delta
^{4}(y-z)\right) .\quad  \label{decomp}
\end{align}%
Here and in what follows, it is understood that after the partial
differentiations all fields are set equal to their constant background
values. Henceforward, these coordinate-independent partial derivatives will
be denoted by subscripts at $\mathfrak{L.}$ For instance $\frac{\partial ^{2}%
\mathfrak{L}(\mathfrak{F}(z),\mathfrak{G}(z))}{\partial \mathfrak{F}%
(z)\partial \mathfrak{G}(z)}=\mathfrak{L}_{\mathfrak{FG}}$. Also, in the
rest of the paper we omit the superscript \emph{ext} by field notations.
Passing to Fourier transforms we identify the scalar coefficients and
matrices in (\ref{lincomb}) as follows\footnote{%
The four matrices (\ref{matrices}) $\Psi _{\mu \tau }^{\left( i\right) }$
may serve as a set of independent transverse matrices, mentioned in the
previous Section in connection with (\ref{lincomb}), in the case of nonlocal
action as well. The following connection with one of the matrices pointed
for that more general case in \cite{batalin}, \cite{Trudy}
\begin{equation*}
\left( Fk\right) _{\mu }\left( F^{3}k\right) _{\nu }+\left( \mu
\leftrightarrow \nu \right) =-\mathfrak{G}\Psi _{\mu \tau }^{\left( 4\right)
}-2\Psi _{\mu \tau }^{\left( 2\right) }
\end{equation*}%
holds}%
\begin{eqnarray}
\Theta _{1} &=&\mathfrak{L}_{\mathfrak{F}},\text{ }\Theta _{2}=\mathfrak{L}_{%
\mathfrak{FF}},\text{ }\Theta _{3}=\mathfrak{L}_{\mathfrak{GG}},\text{ }%
\Theta _{4}=\mathfrak{L}_{\mathfrak{FG}}  \notag \\
\Psi _{\mu \tau }^{\left( 1\right) } &=&\eta _{\mu \tau }k^{2}-k_{\mu
}k_{\tau }  \notag \\
\Psi _{\mu \tau }^{\left( 2\right) } &=&\left( Fk\right) _{\mu }\left(
Fk\right) _{\tau }\text{ \ }\Psi _{\mu \tau }^{\left( 3\right) }=\left(
\widetilde{F}k\right) _{\mu }\left( \widetilde{F}k\right) _{\tau },  \notag
\\
\text{ \ }\Psi _{\mu \tau }^{\left( 4\right) } &=&\left( Fk\right) _{\mu
}\left( \widetilde{F}k\right) _{\tau }+\left( \widetilde{F}k\right) _{\mu
}\left( Fk\right) _{\tau }.  \label{matrices}
\end{eqnarray}%
The second antisymmetric term in (\ref{decomp}) has not contributed, since
it nullifies, when contracted with the symmetric common factor $\left( \frac{%
\partial }{\partial z_{\alpha }}\delta ^{4}(x-z)\right) \left( \frac{%
\partial }{\partial z_{\beta }}\delta ^{4}(y-z)\right) \Longrightarrow
k_{\alpha }k_{\beta }.$ Here designations $\left( Fk\right) _{\mu }=\text{\ }%
F_{\mu \alpha }k^{\alpha }$, $\left( \widetilde{F}k\right) _{\mu }=\text{\ }%
\widetilde{F}_{\mu \alpha }k^{\alpha }$ are adopted. {\Huge \ }

The set (\ref{matrices}) is more ample than the set of tensors exploited in
\cite{batalin}, \cite{Trudy} in the sense that it also covers theories with
spacial parity violation, because the matrix \ $\Psi _{\mu \tau }^{\left(
4\right) }$ is a pseudo-tensor. However, \ it is in parity-even theories
that the event of kinematization of the basis in the degenerate case $%
\mathfrak{G}=$ $0$ occurs, since in these theories Lagrangian $\mathfrak{L}%
\left( \mathfrak{F,G}\right) $ must be a scalar, in other words it must be
an even function of the pseudo-scalar $\mathfrak{G.}$ Then, matrix \ $\Psi
_{\mu \tau }^{\left( 4\right) }$ does not contribute into (\ref{lincomb}) as
it enters with the vanishing coefficient $\left. \mathfrak{L}_{\mathfrak{FG}%
}\right\vert _{\mathfrak{G}=0}$ $=0$; therefore, only three independent
matrices are left.

Apart from $\flat _{\mu }^{(1)}$ (\ref{b1}), two other eigenvectors $\flat
_{\mu }^{(2,3)}$ and eigenvalues $\varkappa _{2,3}$ will be looked for by
using a basis $c_{\mu }^{\pm }$ spanning the two-dimensional subspace
orthogonal to $\flat _{\mu }^{(1)}$ and $k_{\mu },$ $\left( (c^{\pm }\flat
^{(1)})=(c^{\pm }k)=0\right) $\ of two orthonormalized $(c^{+}c^{-})=0,\quad
(c^{\pm })^{2}=1$ vectors . Presenting the two eigenvectors $\flat _{\mu
}^{(2,3)}$ as linear combinations $\flat _{\mu }=a^{+}c_{\mu
}^{+}+a^{-}c_{\mu }^{-}$ we find that the problem (\ref{eigenproblem})
becomes a set of two linear homogeneous equations for the two coefficients $%
a^{\pm }.$ Its solvability condition is given by the quadratic equation with
respect to $\varkappa ,$%
\begin{equation}
\varkappa ^{2}-\varkappa \left( \Pi ^{--}+\Pi ^{++}\right) -\left( \Pi
^{+-}\right) ^{2}+\Pi ^{--}\Pi ^{++}=0,  \label{kappasuar}
\end{equation}%
where \footnote{%
Quantities (\ref{Py+-}) are Lambda's of \cite{ShabUs2010} taken with
opposite sign, see Eq. (13) there.}%
\begin{equation}
\Pi ^{++}=c_{\alpha }^{+}\Pi ^{\alpha \beta }c_{\beta }^{+},\text{ }\Pi
^{+-}=\Pi ^{-+}=c_{\alpha }^{+}\Pi ^{\alpha \beta }c_{\beta }^{-},\text{ }%
\Pi ^{--}=c_{\alpha }^{-}\Pi ^{\alpha \beta }c_{\beta }^{-}\text{ }
\label{Py+-}
\end{equation}%
are scalars that may depend on $\mathfrak{F}$, $\mathfrak{G}$, $k^{2}$ and $%
kF^{2}k\equiv k_{\mu }F^{\mu \nu }F_{\nu \tau }k^{\tau }.$ (The other scalar
$k\widetilde{F}^{2}k\equiv k_{\mu }\widetilde{F}^{\mu \nu }\widetilde{F}%
_{\nu \tau }k^{\tau }$ is not independent due to the relation $\widetilde{F}%
_{\mu \nu }^{2}-F_{\mu \nu }^{2}=2\mathfrak{F}\eta _{\mu \nu }.)$ Because of
\ Eq. (\ref{lincomb}) and taking into account that matrices (\ref{matrices})
are homogeneous functions of second degree of components of the photon
momenta $k_{\mu \text{ }},$ while\ vectors $c_{\mu }^{\pm }$ are normalized
to unity, we may conclude that the matrix elements $\Pi ^{\pm \pm },\Pi
^{\mp \pm \text{ \ }}$(\ref{Py+-}) are homogeneous functions of first degree
of $k^{2}$ and $kF^{2}k.$ This assertion is confirmed by calculated
equations (\ref{PSI}), (\ref{c0mbnations}) below.

Equation (\ref{kappasuar}) has two eigenvalues $\varkappa _{2,3}$ as its
solutions. Instead of solving it, let us substitute the dispersion equation (%
\ref{disp}) into it to obtain a dispersion equation
\begin{equation}
\left( k^{2}\right) ^{2}-k^{2}\left( \Pi ^{--}+\Pi ^{++}\right) -\left( \Pi
^{+-}\right) ^{2}+\Pi ^{--}\Pi ^{++}=0.  \label{components}
\end{equation}%
From now on we restrict our consideration to the special frame, wherein
background electric and magnetic fields are parallel, $\boldsymbol{%
E\parallel B}$ ( $\mathfrak{G>}$ $0$), or antiparallel, $\boldsymbol{%
E\parallel -B}$ ($\mathfrak{G<}$ $0)$\footnote{%
To be more definite, note that the speed of the Lorentz boost taking us to
the special frame from a general frame%
\begin{equation*}
v_{x}^{\pm }=\frac{-\mathbf{E}^{\prime 2}-B_{z}^{\prime 2}\pm \left( \left(
\mathbf{E}^{\prime 2}+B_{z}^{\prime 2}\right) ^{2}-4E_{y}^{\prime
2}B_{z}^{\prime 2}\right) ^{1/2}}{2B_{z}^{\prime }E_{y}^{\prime }}
\end{equation*}%
is obtained by solving a quadratic equation. The both values are real, but
only the one with the upper sign corresponds to a speed smaller than that of
light, $\mid v_{x}^{+}\mid \leqslant 1$, while the other is superluminal.
Here, without loss of generality, we have chosen the boost to be performed
along the axis $x$, while the fields in the general frame (primed one) to
have the only nonvanishing components: $B_{z}^{\prime }>0,$ $E_{y}^{\prime
}\gtrless $ $0,$ $E_{z}^{\prime }\gtrless 0$, so that the invariant $%
\mathfrak{G}$ be admitted to have both signs: $\mathbf{B}^{\prime }\mathbf{E}%
^{\prime }=$ $B_{z}^{\prime }E_{z}^{\prime }\gtrless $ $0.$ Also we denoted $%
\mathbf{E}^{\prime 2}=E_{y}^{\prime 2}+E_{z}^{\prime 2}.$%
\par
After the boost the fields are parallel or antiparallel depending on the
sign of $E_{z}^{\prime },$ but the Lorentz transformation is one and the
same, since the speed $v_{x}^{\pm }$ depends only on the square $%
E_{z}^{\prime 2}.$}. This is sufficient for our purposes, since a signal can
propagate faster than light in no frame. In the special frame one has%
\begin{equation}
\mathfrak{F=}\frac{B^{2}-E^{2}}{2},\text{ \ }\mathfrak{G=}\text{ }\mathbf{BE}%
\text{ }=\mathfrak{\pm }BE,  \label{map}
\end{equation}%
where $B=\mid \mathbf{B\mid },$ $E=\mid \mathbf{E\mid },$ and $%
kF^{2}k=-B^{2}k_{\perp }^{2}+E^{2}\left( k_{\parallel }^{2}-k_{0}^{2}\right)
$, \ where $k_{\parallel }^{2}$ and $\boldsymbol{k}_{\perp }^{2}$ are
squares of photon momentum components, respectively, along and across the
common axis of the background electric and magnetic field. In this frame we
choose the basic vectors $c_{\mu }^{\pm }$ as%
\begin{eqnarray}
c_{\mu }^{-} &=&\frac{{B}(Fk)_{\mu }+{E}(\tilde{F}k)_{\mu }}{({B}^{2}+{E}%
^{2})|\mathbf{k_{\perp }|}},  \notag \\
c_{\mu }^{+} &=&\frac{E(Fk)_{\mu }-B(\tilde{F}k)_{\mu }}{%
(B^{2}+E^{2})(k_{0}^{2}-k_{\parallel }^{2})^{1/2}}.  \label{c+-}
\end{eqnarray}%
To keep the factor $(k_{0}^{2}-k_{\parallel }^{2})^{1/2}$ real and thereby \
$c_{\mu }^{+}c^{+\mu \text{ }}$equal to unity, note that $%
k_{0}^{2}-k_{\parallel }^{2}$ can be done positive by a farther
specialization of the reference frame: namely, the parallelity or
anti-parallelity of electric and magnetic of our hitherto used frame
remains unaffected by an extra Lorentz boost along their common
axis, aimed to
nullifying the momentum component $k_{\parallel }.$ Note that $%
k_{0}^{2}-k_{\parallel }^{2}$ is just the value taken in the special frame
by the invariant $-\frac{k^{2}\mathcal{B}^{2}+kF^{2}k}{\left( \mathfrak{F}%
^{2}+\mathfrak{G}^{2}\right) ^{1/2}},$ where $\mathcal{B}^{2}=\mathfrak{F+}%
\left( \mathfrak{F}^{2}+\mathfrak{G}^{2}\right) ^{1/2}$ . Therefore, once it
is positive in the farther-specialized frame, it remains positive in every
frame.\ Employing the notations
\begin{equation}
x=k_{\perp }^{2},\text{ }\beta x=k^{2}  \label{x,y}
\end{equation}%
(note that if $y=k_{0}^{2}-k_{\parallel }^{2}=\left( 1-\beta \right) x$, \
then $k^{2}=x-y$) we find it useful to present the following intermediate
formulas for certain scalar products\qquad\ \ \ \ \ \ \ \ \ \ \ \ \
\begin{eqnarray}
\left( Fk\right) ^{\mu }c_{\mu }^{+} &=&E\left( \left( 1-\beta \right)
x\right) ^{1/2},\text{ }\ \left( \widetilde{F}k\right) ^{\mu }c_{\mu
}^{+}=-B\left( \left( 1-\beta \right) x\right) ^{1/2},  \notag \\
\left( Fk\right) ^{\mu }c_{\mu }^{-} &=&Bx^{1/2},\text{ \ }\left( \widetilde{%
F}k\right) ^{\mu }c_{\mu }^{-}=Ex^{1/2}  \label{Fkc}
\end{eqnarray}%
and for the matrix elements $\Psi _{++}^{\left( i\right) }=c_{\alpha
}^{+}\Psi ^{\left( i\right) \alpha \beta }c_{\beta }^{+},$ $\Psi
_{+-}^{\left( i\right) }=\Psi _{-+}^{\left( i\right) }=c_{\alpha }^{+}\Psi
^{\left( i\right) \alpha \beta }c_{\beta }^{-},$ $\Psi _{--}^{\left(
i\right) }=c_{\alpha }^{-}\Psi ^{\left( i\right) \alpha \beta }c_{\beta
}^{-} $ \
\begin{eqnarray}
\Psi _{++}^{\left( 1\right) } &=&\beta x,\text{ \ \ \ \ \ \ \ \ \ \ \ \ \ \
\ \ \ \ \ \ \ \ }\Psi _{+-}^{\left( 1\right) }=0,\text{ \ \ \ \ \ \ \ \ \ \
\ \ \ \ \ \ \ \ \ \ \ \ \ \ \ \ \ \ \ \ \ \ \ \ \ \ \ \ }\Psi _{--}^{\left(
1\right) }=\beta x,  \notag \\
\Psi _{++}^{\left( 2\right) } &=&E^{2}\left( 1-\beta \right) x,\text{ \ \ \
\ \ \ \ \ }\Psi _{+-}^{\left( 2\right) }=EB\left( 1-\beta \right) ^{1/2}x,%
\text{\ \ \ \ \ \ \ \ \ \ \ \ \ \ \ \ \ \ }\Psi _{--}^{\left( 2\right)
}=B^{2}x,  \notag \\
\Psi _{++}^{\left( 3\right) } &=&B^{2}\left( 1-\beta \right) x,\text{ \ \ \
\ \ \ }\ \Psi _{+-}^{\left( 3\right) }=-EB\left( 1-\beta \right) ^{1/2}x,%
\text{\ \ \ \ \ \ \ \ \ \ \ \ \ \ \ \ \ \ }\Psi _{--}^{\left( 3\right)
}=E^{2}x,  \notag \\
\Psi _{++}^{\left( 4\right) } &=&-2EB\left( 1-\beta \right) x,\text{ \ \ }%
\Psi _{+-}^{\left( 4\right) }=-\left( B^{2}-E^{2}\right) \left( 1-\beta
\right) ^{1/2}x,\text{\ \ \ }\ \Psi _{--}^{\left( 4\right) }=2EBx
\label{PSI}
\end{eqnarray}%
Finally, for matrix elements $\Pi ^{\pm \pm },\Pi ^{\mp \pm \text{ \ }}$(\ref%
{Py+-}) we have\footnote{%
These coincide with $-\Lambda _{3,2,4}$  given by Eqs. (18) -- (20) of Ref.\cite%
{ShabUs2010} after Eq. (20) is corrected
by substituting $\Gamma _{3}$ for the contents of the bracket.}%
\begin{eqnarray*}
\text{\ }\Pi ^{++} &=&\mathfrak{L}_{\mathfrak{F}}\beta x+x\left( 1-\beta
\right) \Gamma _{2} \\
\Pi ^{--} &=&\mathfrak{L}_{\mathfrak{F}}\beta x+x\Gamma _{1}, \\
\Pi ^{+-} &=&\Gamma _{3}\left( 1-\beta \right) ^{1/2}x,\text{ \ } \\
&&\text{ \ }
\end{eqnarray*}%
where\footnote{%
The Gammas (\ref{Gammas}) are associated with electric permeabilities and
magnetic permittivities -- to be published in a forthcoming paper by T.
Adorno, D. Gitman and the present author.}%
\begin{eqnarray}
\Gamma _{1} &=&\mathfrak{L}_{\mathfrak{FF}}B^{2}+\mathfrak{L}_{\mathfrak{GG}%
}E^{2}+2BE\mathfrak{L}_{\mathfrak{FG}},  \notag \\
\Gamma _{2} &=&\mathfrak{L}_{\mathfrak{FF}}E^{2}+\mathfrak{L}_{\mathfrak{GG}%
}B^{2}-2BE\mathfrak{L}_{\mathfrak{FG}}\mathfrak{,}  \notag \\
\Gamma _{3} &=&\left( \mathfrak{L}_{\mathfrak{FF}}-\mathfrak{L}_{\mathfrak{GG%
}}\right) \mathfrak{G}-\mathfrak{L}_{\mathfrak{FG}}\left( B^{2}-E^{2}\right)
\label{Gammas}
\end{eqnarray}%
and for their combinations involved in (\ref{kappasuar}), (\ref{components})
\begin{equation*}
\Pi ^{++}+\Pi ^{--}=x\left[ 2\beta \mathfrak{L}_{\mathfrak{F}}+\Gamma
_{1}+\Gamma _{2}\left( 1-\beta \right) \right] ,\text{ }
\end{equation*}

\begin{eqnarray}
\Pi ^{++}\Pi ^{--} &=&\left[ \beta ^{2}\mathfrak{L}_{\mathfrak{F}}^{2}+\beta
\mathfrak{L}_{\mathfrak{F}}\left( \Gamma _{1}+\Gamma _{2}\left( 1-\beta
\right) \right) +\Gamma _{1}\Gamma _{2}\left( 1-\beta \right) \right] x^{2},
\notag \\
\text{\ \ \ }\left( \Pi ^{+-}\right) ^{2} &=&\Gamma _{3}^{2}\left( 1-\beta
\right) x^{2}.  \label{c0mbnations}
\end{eqnarray}%
It is notable that Eqs. (\ref{Gammas}) are related to second partial
derivatives of the function $\overline{\mathfrak{L}}\mathfrak{(}B\mathfrak{,}%
E)=\mathfrak{L(}\mathfrak{F}(B\mathfrak{,}E),\mathfrak{G}(B\mathfrak{,}E))$,
obtained from $\mathfrak{L(}\mathfrak{F},\mathfrak{G})$ by the mapping (\ref%
{map}):%
\begin{eqnarray}
\Gamma _{1} &=&\overline{\mathfrak{L}}_{BB}-\mathfrak{L}_{\mathfrak{F}},
\notag \\
\Gamma _{2} &=&\overline{\mathfrak{L}}_{EE}+\mathfrak{L}_{\mathfrak{F}}%
\mathfrak{,}  \notag \\
\Gamma _{3} &=&-\overline{\mathfrak{L}}_{BE}+\mathfrak{L}_{\mathfrak{G}}
\label{parder}
\end{eqnarray}

Now, by substituting (\ref{x,y}) and (\ref{c0mbnations}) into (\ref%
{components}), the latter, outside the origin $x=0$ ( $k_{\perp
}^{2}=k^{2}=0),$ becomes after canceling the factor $x^{2}\neq 0$ the
quadratic equation for $\beta $ (defined by Eq. (\ref{x,y}) as $\beta =\frac{%
k^{2}}{k_{\perp }^{2}}$)
\begin{equation}
\text{ \ \ \ \ \ \ \ \ \ \ \ \ \ \ \ \ }\beta ^{2}\left( 1-\mathfrak{L}_{%
\mathfrak{F}})(1-\mathfrak{L}_{\mathfrak{F}}+\Gamma _{2}\right) -\beta \left[
\left( \Gamma _{1}+\Gamma _{2}\right) (1-\mathfrak{L}_{\mathfrak{F}})+\Gamma
_{1}\Gamma _{2}-\Gamma _{3}^{2}\right] +\Gamma _{1}\Gamma _{2}-\Gamma
_{3}^{2}=0.  \label{eq for beta}
\end{equation}

In understanding that its two solutions $\beta _{2,3}$\ are independent of
photon momenta, Eq. (\ref{eq for beta}) determines two declination angles $%
\arctan (1-\beta _{3,2})$ of the straight lines $y=\left( 1-\beta
_{3,2}\right) x,$ or
\begin{equation}
k_{0}^{2}-k_{\parallel }^{2}=\left( 1-\beta \right) k_{\perp }^{2}
\label{straight lines}
\end{equation}%
in the plane ($y=k_{0}^{2}-k_{\parallel }^{2},$ $x=k_{\perp }^{2}),$
representing two dispersion laws, the birefringence phenomenon.

\subsection{\protect\bigskip Restrictions on group velocity}

\bigskip Following the definition of the group velocity (\ref{2}) we
find that%
\begin{equation*}
\mid \mathbf{v}^{\mathrm{gr}}\mid ^{2}=\frac{k_{\parallel }^{2}+k_{\perp
}^{2}\left( 1-\beta \right) ^{2}}{k_{0}^{2}},
\end{equation*}%
if calculated on dispersion curves (\ref{straight lines}). The causal
restriction $\mid v_{i}^{\mathrm{gr}}\mid ^{2}\leqslant 1$ now takes the
form
\begin{equation*}
k_{0}^{2}-k_{\parallel }^{2}\geqslant \left( 1-\beta \right) ^{2}k_{\perp
}^{2}.
\end{equation*}%
When confronted with (\ref{straight lines}) this results in the following
two conditions prescribed by causality to each of the dispersion curves:%
\begin{equation}
1\geqslant \beta \geqslant 0,  \label{inequal}
\end{equation}%
in other words, both of them should be located in the space-like sector $%
k_{\perp }^{2}\geqslant k_{0}^{2}-k_{\parallel }^{2}\geqslant 0.$

Now we have to turn to quadratic equation (\ref{eq for beta}) to see what
conditions must be obeyed by the Lagrangian derivatives and field invariants
in order that the inequalities (\ref{inequal}) be satisfied by both of its
solutions.

Note, first, that as long as $\varkappa _{2}$ and $\varkappa _{3}$ are,
according to (\ref{eigenproblem}), eigenvalues of the symmetric and\ real
(Hermitian) tensor $\Pi _{\mu \nu }$,%
\begin{equation}
\varkappa _{3,2}=\frac{1}{2}\left( \Pi ^{--}+\Pi ^{++}\pm \left( \mathfrak{D}%
_{1}\right) ^{1/2}\right) =\frac{1}{2}\left( x\left[ 2\beta \mathfrak{L}_{%
\mathfrak{F}}+\Gamma _{1}+\Gamma _{2}\left( 1-\beta \right) \right] \pm
\left( \mathfrak{D}_{1}\right) ^{1/2}\right)  \label{kappa23}
\end{equation}%
these must be both real. (For the sake of continuity to \cite{Shabus2011},
the upper sign relates to index 3 and the lower to 2 here and throughout the
paper)$.$ Correspondingly, positive is the discriminant \
\begin{equation}
\mathfrak{D}_{1}=\left( \Pi ^{++}+\Pi ^{--}\right) ^{2}-4\left( \Pi ^{++}\Pi
^{--}-\left( \Pi ^{+-}\right) ^{2}\right) =x^{2}\left[ \left( \Gamma
_{1}-\left( 1-\beta \right) \text{\ }\Gamma _{2}\right) ^{2}+4\Gamma
_{3}^{2}\left( 1-\beta \right) \right] >0  \label{D1}
\end{equation}%
of the quadratic polynomial in Eq. (\ref{components}) or Eq. (\ref{kappasuar}%
). (The positivity of $\left( 1-\beta \right) $ $=\left(
k_{0}^{2}-k_{\parallel }^{2}\right) \left( k_{\perp }^{2}\right) ^{-1\text{ }%
}$ was commented on under (\ref{c+-})). The discriminant of the quadratic
polynomial in Eq. (\ref{eq for beta}) can be transformed to explicitly
positive form:%
\begin{eqnarray}
\mathfrak{D}_{2} &\mathfrak{=}&\left[ \left( \Gamma _{1}+\Gamma _{2}\right)
(1-\mathfrak{L}_{\mathfrak{F}})+\Gamma _{1}\Gamma _{2}-\Gamma _{3}^{2}\right]
^{2}-4\left( 1-\mathfrak{L}_{\mathfrak{F}})(1-\mathfrak{L}_{\mathfrak{F}%
}+\Gamma _{2}\right) \left( \Gamma _{1}\Gamma _{2}-\Gamma _{3}^{2}\right) =
\notag \\
&=&\left[ \left( \Gamma _{1}-\Gamma _{2}\right) (1-\mathfrak{L}_{\mathfrak{F}%
})+\Gamma _{1}\Gamma _{2}-\Gamma _{3}^{2}\right] ^{2}+4\Gamma _{3}^{2}(1-%
\mathfrak{L}_{\mathfrak{F}})^{2}>0.  \label{D2}
\end{eqnarray}%
So, both roots of Eq. (\ref{eq for beta}) are real:%
\begin{equation}
\beta _{3,2}=\frac{\left( \Gamma _{1}+\Gamma _{2}\right) (1-\mathfrak{L}_{%
\mathfrak{F}})-\Gamma _{3}^{2}+\Gamma _{1}\Gamma _{2}\pm \mathfrak{D}%
_{2}^{1/2}}{2\left( 1-\mathfrak{L}_{\mathfrak{F}})(1-\mathfrak{L}_{\mathfrak{%
F}}+\Gamma _{2}\right) }.  \label{beta2,3}
\end{equation}%
Note, that the absence of birefringence, $\beta _{3}=\beta _{2},$ $\mathfrak{%
D}_{2}=0$ is only possible provided $\mathfrak{G=}$ $\Gamma _{3}=0$ and $%
\left( \Gamma _{1}-\Gamma _{2}\right) (1-\mathfrak{L}_{\mathfrak{F}})+\Gamma
_{1}\Gamma _{2}=0.$ This is the case \cite{Jerzy} of Born-Infeld Lagrangian.
Complexity of roots $\beta _{3,2}$ would imply instability of the background
-- the constant electromagnetic field -- under growing excitations (in spite
of the fact that the nonlinear Lagrangian has been taken real). It follows
from (\ref{D2}) that no such instability takes place at least within the
present context of local action irrespective of the causality.

In the degenerate case of a parity-even theory with $\mathfrak{G}=0$, $%
\mathfrak{F}>0$, where there is solely a magnetic field in the special
frame, $E=0,$ one has $\mathfrak{L}_{\mathfrak{FG}}^{\left( \mathfrak{G}%
=0\right) }=0$ (since $\mathfrak{L}$ may depend on the pseudoscalar $%
\mathfrak{G}$ only in an even way), hence$\ \Gamma _{3}=0$ and discriminant $%
\mathfrak{D}_{2}$ (\ref{D2})\ becomes a full square (as well as discriminant
$\mathfrak{D}_{1}$ (\ref{D1})). Correspondingly, the roots $\beta
_{2,3}^{\left( \mathfrak{G}=0\right) }$ are:%
\begin{equation}
\beta _{3}^{\left( \mathfrak{G}=0\right) }=\left. \frac{\Gamma _{1}}{1-%
\mathfrak{L}_{\mathfrak{F}}}\right\vert _{E=0}=\frac{\mathfrak{L}_{\mathfrak{%
FF}}B^{2}}{1-\mathfrak{L}_{\mathfrak{F}}},\text{ and }\beta _{2}^{\left(
\mathfrak{G}=0\right) }=\left. \frac{\Gamma _{2}}{1-\mathfrak{L}_{\mathfrak{F%
}}+\Gamma _{2}}\right\vert _{E=0}=\frac{\mathfrak{L}_{\mathfrak{GG}}B^{2}}{1-%
\mathfrak{L}_{\mathfrak{F}}+\mathfrak{L}_{\mathfrak{GG}}B^{2}}\text{ }
\label{betaG=0}
\end{equation}%
for the upper and lower sign in (\ref{beta2,3}), respectively. With these
values, dispersion curves (\ref{straight lines}) reproduce the ones
corresponding to Eqs. (25), (26), respectively, of \ Ref. \cite{Shabus2011}
, where the case of $\mathfrak{G}=0$, $\mathfrak{F}>0$ was studied.

\ Now, to demand that the two conditions $\beta _{3,2}\geqslant 0$ be
fulfilled simultaneously is the same as to demand the fulfilment of the
couple of two conditions:
\begin{equation}
\beta _{2}\beta _{3}\geqslant 0,\ \ \beta _{2}+\beta _{3}\geqslant 0.
\label{c0upl}
\end{equation}%
We have to introduce the additional principle of "Maxwell dominance" at this
step\ in the form%
\begin{equation}
\left( 1-\mathfrak{L}_{\mathfrak{F}})>0,\text{ \ }(1-\mathfrak{L}_{\mathfrak{%
F}}+\Gamma _{2}\right) >0  \label{d0mnanc}
\end{equation}%
claiming that the nonlinearity should not be too strong, so that neither of
the nonlinear contributions $\mathfrak{L}_{\mathfrak{F}},$ $\mathfrak{L}_{%
\mathfrak{F}}-\Gamma _{2}$ in the denominator of (\ref{beta2,3}), might
overwhelm the unity to make the result negative. It will be demonstrated in
the next Section that, in the degenerate mono-field case $\mathfrak{G=}$ $%
\Gamma _{3}=0,$ the inequality (\ref{d0mnanc}) follows literally from the
requirement that poles of the photon propagator on the mass-shells should
have positive residues.

Following Vieta's theorem the first condition in (\ref{c0upl}) is
\begin{equation}
\beta _{2}\beta _{3}=\frac{\Gamma _{2}\Gamma _{1}-\Gamma _{3}^{2}}{\left( 1-%
\mathfrak{L}_{\mathfrak{F}})(1-\mathfrak{L}_{\mathfrak{F}}+\Gamma
_{2}\right) }\geqslant 0.  \label{viete1}
\end{equation}%
Then in view of (\ref{d0mnanc}) it becomes $\Gamma _{2}\Gamma _{1}-\Gamma
_{3}^{2}\geqslant 0.$ Using (\ref{Gammas}) one can find that $\Gamma
_{2}\Gamma _{1}-\Gamma _{3}^{2}=\left( E^{2}+B^{2}\right) ^{2}\left(
\mathfrak{L}_{\mathfrak{FF}}\mathfrak{L}_{\mathfrak{GG}}-\mathfrak{L}_{%
\mathfrak{FG}}^{2}\right) .$ Therefore, we are left with the claimed
condition of nonnegativity of the Hessian
\begin{equation}
\mathfrak{L}_{\mathfrak{FF}}\mathfrak{L}_{\mathfrak{GG}}-\mathfrak{L}_{%
\mathfrak{FG}}^{2}\geqslant 0.  \label{Hessian2}
\end{equation}%
The second condition in (\ref{c0upl}) in accord with the other Vieta's
theorem is
\begin{eqnarray}
\beta _{2}+\beta _{3} &=&\frac{\left( \Gamma _{1}+\Gamma _{2}\right) (1-%
\mathfrak{L}_{\mathfrak{F}})-\Gamma _{3}^{2}+\Gamma _{1}\Gamma _{2}}{\left(
1-\mathfrak{L}_{\mathfrak{F}})(1-\mathfrak{L}_{\mathfrak{F}}+\Gamma
_{2}\right) }=  \notag \\
&=&\frac{\Gamma _{1}}{1-\mathfrak{L}_{\mathfrak{F}}}+\frac{\Gamma _{2}}{1-%
\mathfrak{L}_{\mathfrak{F}}+\Gamma _{2}}-\frac{\Gamma _{3}^{2}}{\left( 1-%
\mathfrak{L}_{\mathfrak{F}})(1-\mathfrak{L}_{\mathfrak{F}}+\Gamma
_{2}\right) }\geqslant 0.  \label{viete2}
\end{eqnarray}%
According to the already established condition (\ref{viete1}) and referring
to (\ref{d0mnanc}) again we are to claim that the product $\Gamma _{2}\Gamma
_{1}$ is positive, hence $\Gamma _{2}$ and $\Gamma _{1}$ are either both
positive or both negative. Now it follows from (\ref{viete2}) that the only
possibility out of \ these two is that the both are positive. So, for
fulfillment of causality it is necessary that
\begin{equation}
\Gamma _{1}>0,\text{ \ }\Gamma _{2}>0.  \label{gamma1,2>0}
\end{equation}%
Then, condition of causality (\ref{viete2}) is guarantied, since the
numerator in its first line is a sum of two positive numbers. As Eq. (\ref%
{gamma1,2>0}) implies also that $\Gamma _{1}+\Gamma _{2}>0$ we see from Eqs.
(\ref{Gammas}) that one of necessary conditions of causality is
\begin{equation}
\mathfrak{L}_{\mathfrak{GG}}+\mathfrak{L}_{\mathfrak{FF}}\geqslant 0.
\label{though}
\end{equation}%
It follows from (\ref{Hessian2}), that $\mathfrak{L}_{\mathfrak{FF}}%
\mathfrak{L}_{\mathfrak{GG}}>0,$ then, taking into account (\ref{though}),
we have
\begin{equation}
\mathfrak{L}_{\mathfrak{GG}}>0,\text{ \ }\mathfrak{L}_{\mathfrak{FF}}>0.
\label{c0nfrms}
\end{equation}
This reinstates, for the general case $\mathfrak{G}\neq 0,$\ conditions
obtained earlier \cite{convexity}, \cite{Shabus2011} for the degenerate
mono-field case $\mathfrak{G}=0.$ Conditions (\ref{Hessian2}) and (\ref%
{c0nfrms}) do not contain fields explicitly and together these make
an extension of our earlier result, that had the same form as
(\ref{c0nfrms}), but with $\mathfrak{G}=0.$ Eqs. (\ref{gamma1,2>0})
contain additional, field-dependent information.

Remind that Eqs. (\ref{viete2}), (\ref{gamma1,2>0}) relate to the special
reference frame. To get their invariant form one may substitute $\mathfrak{F+%
}\sqrt{\mathfrak{F}^{2}+\mathfrak{G}^{2}}$ for $B^{2}$, and $-\mathfrak{F+}%
\sqrt{\mathfrak{F}^{2}+\mathfrak{G}^{2}}$ for $E^{2}$ in Eqs. (\ref{Gammas})
to get $\Gamma _{1,2,3}$ in terms of field invariants and derivatives with
respect to them.

We conclude that the right-hand inequality in the causality condition (\ref%
{inequal}) for two roots is equivalent -- via the principle of "Maxwell
dominance" -- to $\ $two conditions (\ref{Hessian2}) and (\ref{gamma1,2>0}),
of which the first one is of pure geometric nature and relates only to the
properties of function $\mathfrak{L(F,G),}$while the second may be a
restriction on the strength of the fields. The reservation is in order,
though, that condition (\ref{though}) also is field-independent like (\ref%
{Hessian2}), but the geometrical condition (\ref{though}) is weaker than (%
\ref{gamma1,2>0}).

As for the left inequality in (\ref{inequal}), the two corresponding
relations $\beta _{3,2}\leqslant 1$ for the roots (\ref{beta2,3}) of the
quadratic trinomial in Eq.(\ref{eq for beta})) are both satisfied already
due to (another manifestation of) the same "Maxwell dominance" principle
written in the form
\begin{equation}
\left( \Gamma _{1}+\Gamma _{2}\right) (1-\mathfrak{L}_{\mathfrak{F}})-\Gamma
_{3}^{2}+\Gamma _{1}\Gamma _{2}\pm \mathfrak{D}_{2}^{1/2}<2\left( 1-%
\mathfrak{L}_{\mathfrak{F}})(1-\mathfrak{L}_{\mathfrak{F}}+\Gamma
_{2}\right) .  \label{doominance2}
\end{equation}%
According to this principle, the unity in the right-hand part dominates over
all nonlinear, i.e. containing $\Gamma $'s (\ref{Gammas}) and $\mathfrak{L}_{%
\mathfrak{F}},$ terms and thus provides \ fulfillment of the inequality (\ref%
{doominance2}).

\subsection{\protect\bigskip Correspondence principle}

It remains to impose the requirement of correspondence with the linear
(Maxwell) electrodynamics in the limit of vanishing field. It rules that the
nonlinear correction to the full Lagrangian $L$ (\ref{S}) should disappear
together with its first derivatives in this limit:%
\begin{equation}
\left. \mathfrak{L}\right\vert _{\mathfrak{F=G}=0}=0,\left. \frac{\partial
\mathfrak{L}}{\partial \mathfrak{F}}\right\vert _{\mathfrak{F=G}=0}=0,\
\left. \frac{\partial \mathfrak{L}}{\partial \mathfrak{G}}\right\vert _{%
\mathfrak{F=G}=0}=0.  \label{correspondence}
\end{equation}%
While the first of these relations means that the surface given by the
function $Z=$ $\mathfrak{L}\left( \mathfrak{F,G}\right) $ -- whose curvature
has been proven above to be positive for all fields below the limits
admitted by the Maxwell dominance principle -- contains the point of the
origin $\mathfrak{F=G}=0,$ the other two imply that this surface is tangent
to the coordinate plane $Z=0$ in the origin, the latter being its point of
extremum.

According to (\ref{beta2,3}) it holds that $\mathfrak{L}_{\mathfrak{FF}%
}\geqslant 0,$ $\mathfrak{L}_{\mathfrak{GG}}\geqslant 0$ everywhere, the
origin $\mathfrak{F=G}=0$ included$.$ With the account of this fact let us
prove that this extremum is a maximum. Consider a family of curves,
parameterized by parameter $\theta ,$ which is formed by the crossing of the
surface $Z=$ $\mathfrak{L}\left( \mathfrak{F,G}\right) $ with the plane
given in the 3-space ($Z,\mathfrak{F,G)}$ by equation $\mathfrak{G=\theta F.}
$We have to prove that along each curve the full second derivative in the
origin $\mathfrak{F=G}=0$%
\begin{equation}
\frac{d^{2}\mathfrak{L}\left( \mathfrak{F,\theta F}\right) }{d\mathfrak{F}%
^{2}}=\mathfrak{L}_{\mathfrak{FF}}+2\theta \mathfrak{L}_{\mathfrak{FG}%
}+\theta ^{2}\mathfrak{L}_{\mathfrak{GG}}\geqslant 0  \label{fullder}
\end{equation}%
is positive. This fact is immediately evident in any parity-even theory
because $\mathfrak{L}_{\mathfrak{FG}}=0$ at $\mathfrak{G=}$ $0$ in such
theory. On the contrary, in a theory with parity violation $\mathfrak{L}$
may include odd powers of the pseudoscalar $\mathfrak{G,}$but the inequality
(\ref{fullder}) still holds true. Note, first, that (\ref{fullder}) is
obviously valid at $\theta =0$ due to (\ref{beta2,3}). If (\ref{fullder})
might be violated at another value of $\theta ,$ the origin $\mathfrak{F=G}%
=0 $ would be a saddle point of the surface given by the function $Z=$ $%
\mathfrak{L}\left( \mathfrak{F,G}\right) $. This is, however, forbidden by
condition (\ref{Hessian2}) that states that the curvature is everywhere
positive. Let us see this directly. Let, first, $\theta $ be positive. Then
Eq. (\ref{fullder}) can only be violated provided $\mathfrak{L}_{\mathfrak{FG%
}}$ is negative at $\mathfrak{G=}$ $0.$ Eq. (\ref{Hessian2}) leaves for $%
\mathfrak{L}_{\mathfrak{FG}}$ the only possibility $\mathfrak{L}_{\mathfrak{%
FG}}>-$ $\left( \mathfrak{L}_{\mathfrak{FF}}\mathfrak{L}_{\mathfrak{FG}%
}\right) ^{1/2}.$ Then, for the full derivative in (\ref{fullder}) we can
write\bigskip
\begin{eqnarray}
\frac{d^{2}\mathfrak{L}\left( \mathfrak{F,\theta F}\right) }{d\mathfrak{F}%
^{2}} &=&\mathfrak{L}_{\mathfrak{FF}}+2\theta \mathfrak{L}_{\mathfrak{FG}%
}+\theta ^{2}\mathfrak{L}_{\mathfrak{GG}}>\mathfrak{L}_{\mathfrak{FF}%
}-2\theta \left( \mathfrak{L}_{\mathfrak{FF}}\mathfrak{L}_{\mathfrak{FG}%
}\right) ^{1/2}+\theta ^{2}\mathfrak{L}_{\mathfrak{GG}}= \\
&=&\left( \mathfrak{L}_{\mathfrak{FF}}^{1/2}-\theta \mathfrak{L}_{\mathfrak{%
GG}}^{1/2}\right) ^{2}>0.
\end{eqnarray}%
Analogously, when $\theta $ is negative (\ref{fullder}) can only be violated
provided $\mathfrak{L}_{\mathfrak{FG}}$ is positive. Eq. (\ref{Hessian2})
leaves for $\mathfrak{L}_{\mathfrak{FG}}$ the only possibility $\mathfrak{L}%
_{\mathfrak{FG}}<$ $\left( \mathfrak{L}_{\mathfrak{FF}}\mathfrak{L}_{%
\mathfrak{FG}}\right) ^{1/2}.$ Then, for the full derivative in (\ref%
{fullder}) we can write\bigskip
\begin{eqnarray}
\frac{d^{2}\mathfrak{L}\left( \mathfrak{F,\theta F}\right) }{d\mathfrak{F}%
^{2}} &=&\mathfrak{L}_{\mathfrak{FF}}+2\theta \mathfrak{L}_{\mathfrak{FG}%
}+\theta ^{2}\mathfrak{L}_{\mathfrak{GG}}>\mathfrak{L}_{\mathfrak{FF}%
}+2\theta \left( \mathfrak{L}_{\mathfrak{FF}}\mathfrak{L}_{\mathfrak{FG}%
}\right) ^{1/2}+\theta ^{2}\mathfrak{L}_{\mathfrak{GG}}= \\
&=&\left( \mathfrak{L}_{\mathfrak{FF}}^{1/2}+\theta \mathfrak{L}_{\mathfrak{%
GG}}^{1/2}\right) ^{2}>0.
\end{eqnarray}%
So, we have verified that condition (\ref{fullder}) can be violated in
neither case, there is no descent along any direction $\arctan \theta $ in
the $\left( \mathfrak{F,G}\right) $-plane from the point $\mathfrak{F=G}=0$,
the latter being neither a maximum, nor a saddle, but a minimum point of the
surface given by the function $Z=$ $\mathfrak{L}\left( \mathfrak{F,G}\right)
$ in theories both with and without parity violation.

Turning now to Eq. (\ref{parder}) we see that in the extremum point $%
\mathfrak{L}_{\mathfrak{F}}=\mathfrak{L}_{\mathfrak{G}}=0,$ condition (\ref%
{Hessian2}) becomes $\overline{\mathfrak{L}}_{BB}\overline{\mathfrak{L}}%
_{EE}\geqslant \overline{\mathfrak{L}}_{BE}^{2}$, which implies the
convexity of the surface given by the Lagrangian $\overline{\mathfrak{L}}%
(B,E) $ mapped to the space of variables $B,E,$ which are the fields in the
special reference frame, at least in that point. It seems plausible that
this statement may be extended to other points of that surface, but this
requires additional consideration.

\section{\protect\bigskip Photon Green function, its poles and residues}

The Green function of a photon $D_{\mu \nu }(k)$ is defined as a solution of
the equation obtained from Eq. (\ref{algeb}) by replacing zero in the
right-hand side by unity
\begin{equation}
\left[ \eta _{\mu \nu }k^{2}-k_{\mu }k_{\nu }-\Pi _{\mu \nu }(k)\right]
D^{\nu \tau }(k)=\eta _{\mu }^{\tau }.  \label{qgreen}
\end{equation}%
That equation is solved by the formula\cite{batalin}
\begin{equation}
D_{\mu \nu }(k)=\sum_{c=1}^{3}\frac{\flat _{\mu }^{(c)}~\flat _{\nu }^{(c)}}{%
(\flat ^{(c)})^{2}}\frac{1}{k^{2}-\varkappa _{c}(k)}  \label{Green}
\end{equation}%
(up to a transverse term $k_{\mu }k_{\nu }$ that remains arbitrary). This
fact is verified by direct substitution of (\ref{Green}) into (\ref{qgreen})
with the use of (\ref{diag}) and of orthogonality of eigenvectors $\flat
_{\mu }^{(c)},$ and of their completeness $\sum_{c=1}^{3}\frac{\flat _{\mu
}^{(c)}~\flat _{\tau }^{(c)}}{(\flat ^{(c)})^{2}}=\eta _{\mu }^{\tau }-\frac{%
k_{\mu }k_{\tau }}{k^{2}}$ in the 3-dimensional subspace, transversal to $%
k_{\mu }.$

In this Section we consider consequences of the requirement that residues of
the Green function (\ref{Green}) in poles $k^{2}-\varkappa _{c}(k)$ \
corresponding to each eigenmode $c=1,2,3$ be positive.

Using the form (\ref{b1}) of the eigenvector $\flat _{\mu }^{(1)}$ of the
first (nonpropagating) mode and the representation (\ref{matrices}) we find
for its eigenvalue $\varkappa _{1}(k)=\mathfrak{L}_{\mathfrak{F}}k^{2}.$
Hence the residue in the first term in (\ref{Green}) is positive when
\begin{equation}
1-\mathfrak{L}_{\mathfrak{F}}>0  \label{R}
\end{equation}%
which is within the range of the Maxwell-dominance principle.

\bigskip Consider now any of the modes 2 or 3.~The residue of the pole in (%
\ref{Green}) is positive when%
\begin{equation}
\frac{d}{dk^{2}}\left( k^{2}-\varkappa _{3,2}(k)\right) \left. \text{ }%
\right\vert _{k^{2}=\varkappa _{_{3,2}}(k)}=1-\frac{1}{x}\frac{d\varkappa
_{3,2}}{d\beta }>0.  \label{rsduy}
\end{equation}%
Here $\varkappa _{2,3}$ is any of the two solutions of equation (\ref%
{kappasuar})%
\begin{equation*}
\varkappa _{3,2}=\frac{1}{2}\left[ x\left( 2\mathfrak{L}_{\mathfrak{F}}\beta
+\Gamma _{1}+\left( 1-\beta \right) \text{\ }\Gamma _{2}\right) \pm
\mathfrak{D}_{1}^{1/2}\right] ,
\end{equation*}%
the discriminant $\mathfrak{D}_{1}$ being given by (\ref{D1}). Now condition
(\ref{rsduy}) takes the form%
\begin{equation}
1-\frac{1}{x}\frac{d\varkappa _{3,2}}{d\beta }=1-\mathfrak{L}_{\mathfrak{F}}+%
\frac{\Gamma _{2}}{2}\mp \frac{\Gamma _{1}\Gamma _{2}-\Gamma _{2}^{2}\left(
1-\beta \right) -2\Gamma _{3}^{2}}{2\left[ \left( \Gamma _{1}-\Gamma
_{2}\left( 1-\beta \right) \right) ^{2}+4\Gamma _{3}^{2}\left( 1-\beta
\right) \right] ^{1/2}}>0.  \label{<ref>rsdsp0s</ref>}
\end{equation}%
Its fulfillment is provided within the domain of Maxwell dominance.

\bigskip In the degenerate mono-field case, $\Gamma _{3}=0,$ residue (\ref%
{<ref>rsdsp0s</ref>}) reduces to%
\begin{eqnarray}
\text{Res }D_{3,2} &=&1-\frac{1}{x}\frac{d\varkappa _{3,2}}{d\beta }=1-%
\mathfrak{L}_{\mathfrak{F}}+\frac{\Gamma _{2}}{2}\mp \frac{\Gamma _{1}\Gamma
_{2}-\Gamma _{2}^{2}\left( 1-\beta \right) }{2\mid \Gamma _{1}-\Gamma
_{2}\left( 1-\beta \right) ^{2}\mid }=  \notag \\
&=&1-\mathfrak{L}_{\mathfrak{F}}+\frac{\Gamma _{2}}{2}\mp \frac{\Gamma _{2}}{%
2}\text{sgn}\left( \Gamma _{1}-\Gamma _{2}\left( 1-\beta \right) \right) .
\label{residuesG=0}
\end{eqnarray}%
Therefore,%
\begin{equation*}
\text{Res }D_{3,2}=\text{ \ }\left\{
\begin{tabular}{cc}
\ $1-\mathfrak{L}_{\mathfrak{F}}$ &  \\
$1-\mathfrak{L}_{\mathfrak{F}}+\Gamma _{2}$ & ~%
\end{tabular}%
\text{if \ \ }\left( \Gamma _{1}-\Gamma _{2}\left( 1-\beta \right) \right)
>0,\right.
\end{equation*}%
and%
\begin{equation*}
\text{Res }D_{3,2}=\text{ \ }\left\{
\begin{tabular}{cc}
\ $1-\mathfrak{L}_{\mathfrak{F}}+\Gamma _{2}$ &  \\
$1-\mathfrak{L}_{\mathfrak{F}}$ & ~%
\end{tabular}%
\text{if \ \ }\left( \Gamma _{1}-\Gamma _{2}\left( 1-\beta \right) \right)
<0.\right.
\end{equation*}%
In either case the requirement of positivity of residues of the Green
function in \ all mass-shell poles is that%
\begin{equation}
\left( 1-\mathfrak{L}_{\mathfrak{F}}\right) >0,\text{ }\left( 1-\mathfrak{L}%
_{\mathfrak{F}}+\Gamma _{2}\right) >0  \label{simultaneous}
\end{equation}%
simultaneously. As for the factor \ $\left( \Gamma _{1}-\Gamma _{2}\left(
1-\beta \right) \right) $, it is equal
\begin{equation*}
\text{to \ }\frac{\left( \Gamma _{1}-\Gamma _{2}\right) \left( 1-\mathfrak{L}%
_{\mathfrak{F}}\right) -\Gamma _{1}\Gamma _{2}}{\left( 1-\mathfrak{L}_{%
\mathfrak{F}}\right) }\text{ for }\beta =\beta _{3},\text{ and to }\frac{%
\left( \Gamma _{1}-\Gamma _{2}\right) \left( 1-\mathfrak{L}_{\mathfrak{F}%
}\right) -\Gamma _{1}\Gamma _{2}}{\left( 1-\mathfrak{L}_{\mathfrak{F}%
}+\Gamma _{2}\right) }\text{ for }\beta =\beta _{2}.
\end{equation*}%
In view of (\ref{simultaneous}) we have sgn$\left( \Gamma _{1}-\Gamma
_{2}\left( 1-\beta \right) \right) =$ sgn$\left[ \left( \Gamma _{1}-\Gamma
_{2}\right) \left( 1-\mathfrak{L}_{\mathfrak{F}}\right) -\Gamma _{1}\Gamma
_{2}\right] .$ Interestingly, this factor turns to zero under the
substitution $\beta =\beta _{3,2}$\ in the case where the birefringence is
away, $\beta _{2}=\beta _{3},$ which takes place only for Born-Infeld
Lagrangian according to \cite{Jerzy}. Requirements (\ref{simultaneous})
coincide with Eqs. (27), (28) from \cite{Shabus2011}, bearing in mind that $%
\Gamma _{2}=\mathfrak{L}_{\mathfrak{GG}}B^{2}$ in the degenerate case.

\section{ Conclusion}

In this article we inferred geometrical and other consequences to be obeyed
by local electromagnetic nonlinear effective Lagrangian that follow from the
causality principle understood as a ban for the group velocity of a
long-wave and slow excitations over the constant external field to exceed
the speed of light in the vacuum -- call it just Group Velocity Conditions
(GVC) -- combined with another, \ Maxwell \ dominance, principle \
prescribing that the effects of nonlinearity may correct, but should not
drastically outweigh that which is determined by the linear classical
electrodynamics of \ Faraday and Maxwell. In fact, Maxwell \ dominance
principle (MDP) implies that the corresponding model should contain at least
one small parameter such that the Maxwell electrodynamics be reproduced when
it tends to zero. Specifically, MDP shows itself in the form of three
inequalities, (42), (49) and (60), wherein the nonlinearity-containing terms
are not permitted to change the positive sign fixed by the unity. Eq. (59)
turns out to be equivalent to imposing the requirement of positivity of
residues of poles of the photon propagator on mass shells of photon modes--
call it Positive Residue Conditions (PRC). At the same time, MDP, realized
as Eq. (49), gives the result, coinciding with one of the consequences of
GVC. Namely, the one stating that the both coefficients $\beta _{3,2},$
giving the slopes of two photon dispersion curves, which are as a matter of
fact straight lines in the momentum plane ($k_{0}^{2}-k_{\parallel }^{2},$ $%
k_{\perp }^{2}),$ are less, than unity, $\beta _{3,2}\leqslant 1.$ It is
more remarkable, that MDP in the form of Eq. (42), when combined with the
other consequence of GVC, $\beta _{3,2}\geqslant 0,$ leads to a few
obligatory relations, Eqs.(\ref{Hessian2}) and (\ref{gamma1,2>0}), (\ref%
{c0nfrms}), to be obeyed by derivatives of the local effective Lagrangian
with respect to field invariants. The said is illustrated by the scheme
below.

$\fbox{Curvature, Eqs.(44) and (46), (48)}$ \ \ ${\Huge \Longleftarrow
\impliedby }$ \ $\left\{
\begin{tabular}{cc}
\ Maxwell \ dominance, Eq.(42) &  \\
Causality $\left( \text{GVC}\right) $ $\beta \geqslant 0$, Eq.(35) & ~%
\end{tabular}%
\;.\right. $

$\fbox{Causality $\left( \text{GVC}\right) $\ $\beta \leqslant 1$, \ Eq.(35)
}$ $%
\begin{tabular}{cc}
$\ \ {\Huge \Longleftarrow \impliedby }$\ \ \ \ Maxwell \ dominance, Eq.(49)
&
\end{tabular}%
$

$\fbox{Positive\ residues}$ \ \ \ \ \ \ \ \ \ \ \ \ \ \ $\ \ \ \ \ \ \ \ \ \
\ \ {\Huge \Longleftarrow \impliedby }$ \ \ Maxwell\ dominance, Eq.(60)\ \

\ \ \ Eqs.(44) and (46), (48) may be thought of as the basic result of the
present work. It states the positivity of the curvature of the
two-dimensional \ surface given by the nonlinear local Lagrangian as a
function of two field invariants $\mathfrak{F}$ and $\mathfrak{G}$. Then,
imposing the correspondence principle, Eq. (\ref{correspondence}), we
ascertain that the origin $\mathfrak{F}=$ $\mathfrak{G=}$ $0$\ is the
minimum of the function $\mathfrak{L}\left( \mathfrak{F},\mathfrak{G}\right)
.$ By using, instead, the two variables $E$ and $B,$ which are the values
taken by electric and magnetic fields of the background in the special
Lorentz frame, where these fields are (anti)parallel, we also establish the
convexity of the surface given by the Lagrangian in these variables in the
origin $E=B=0.$

\bigskip

\bigskip

\section{Appendix. \\ Energy conditions versus causality and
positivity of residues \ }

Consequences of causality as imposed on nonlinear effective Lagrangian\ may
be believed to be infinitely numerous, because no sort of a signal is
permitted to travel faster than light. So, by considering different signals
and different circumstances of their propagation one can create more and
more necessary causality conditions, none of them being thought of as
sufficient. For instance, studying propagation of shorter waves against a
constant background should result in certain conditions to be imposed on
nonlocal nonlinear Lagrangian.

Generally accepted are alternative conditions, akin to causality --
so-called postulates of Weak and Dominant Energy Conditions (WEC and DEC)
\cite{ellis}. In a conformal-invariant nonlinear theory fulfillment of these
postulates was considered in \cite{Denhsov}, \cite{sokolov2}. In the
electromagnetic signal propagation problem considered in (at least)
Minkowskian space-time WEC and DEC reduce to the statements, respectively,
that the energy-momentum density should be positive $t_{00}\geqslant 0$ and
its\ flux $t_{0\nu \text{ }}$should be time-like: $t_{0\nu \text{ }%
}^{2}\leqslant 0$ . (The other requirements called Strong Energy
Conditions (SEC) reduce to DEC, since the\ energy-momentum tensor is
traceless in the case under consideration.) The energy conditions
establish "normal" energy-pressure relations of matter, whose
violation may result, for instance, in the loss of attractiveness
property of gravity \cite{wald}, when such matter is used as a
source in Einstein equations determining the metrics of space-time.
We do not touch the question about general consequences concerning
the causal structure of the space-time manifold created by Einstein
equations with matter not subjected to WEC and DEC, especially
exclusion of closed time-cycles and influences beyond the light
cone.

Instead, in this Appendix we perform a perfected -- as confronted to \cite%
{convexity}, \cite{Shabus2011} -- comparison between the impacts on the
nonlinear Lagrangian provided by applying WEC and DEC, on one hand, and our
GVC and PRC, on the other, based on results of our previous works \cite%
{convexity}, \cite{Shabus2011} (reproduced as a special limit in the present
work), where propagation against the background of a constant field with $%
\mathfrak{G}=0$ was considered.

Our first conclusion deduced is that the principle called WEC -- weak energy
condition -- that claims positivity of the energy density $t_{00}\geqslant 0$
\ -- is equivalent, within our problem of photon propagation, to PRC.
Indeed, for modes 3 and 2 ( see Eqs. (56), (59) in \cite{Shabus2011}) WECs
reduce to%
\begin{equation}
\text{WEC}_{\text{3}}\text{: \ \ }1-\mathfrak{L}_{\mathfrak{F}}\geqslant 0
\label{1-Lf}
\end{equation}%
and to
\begin{equation}
\text{ WEC}_{\text{2}}\text{: \ \ }1-\mathfrak{L}_{\mathfrak{F}}+\mathfrak{L}%
_{\mathfrak{GG}}B^{2}\geqslant 0,  \label{!+LGG}
\end{equation}%
Both are the same as conditions (\ref{residuesG=0})\ of positivity of
residues $\left( \text{PRC}\right) $\ of the photon propagator in the poles
corresponding to mass shells of modes 3, 2. Thus, effectively WEC = PRC.
Conditions (\ref{1-Lf}) , (\ref{!+LGG}) lay within the scope of Maxwell
dominance principle, but we stress that we are not using this principle as
an independent source of conditions when dealing with the case $\mathfrak{G=}
$ $0$.

Our second conclusion deduced is that the causality requirements following
from GVC are more restrictive than those of DEC, which is somewhat
unexpected. Indeed, DEC requires that \footnote{%
Within this Appendix all field-derivatives, as $\mathfrak{L}_{\mathfrak{FF}},%
\mathfrak{L}_{\mathfrak{GG}},\mathfrak{L}_{\mathfrak{F}}$\ are meant to be
taken with $\mathfrak{G}$ set equal to zero after differentiation.}$_{\text{
}}$%
\begin{equation}
\text{DEC}_{\text{2}}\text{: \ \ }t_{0\nu \text{ }}^{(2)2}=-\mathfrak{L}_{%
\mathfrak{GG}}B^{2}\left( 1-\mathfrak{L}_{\mathfrak{F}}+\mathfrak{L}_{%
\mathfrak{GG}}B^{2}\right) k_{0}^{2}\left( k_{0}^{2}-k_{\parallel
}^{2}\right) \leqslant 0  \label{Poynting2}
\end{equation}
for mode 2, and that$_{\text{ }}$%
\begin{equation}
\text{DEC}_{\text{3}}\text{: \ \ \ \ \ \ }t_{0\nu \text{ }}^{(3)2}=-%
\mathfrak{L}_{\mathfrak{FF}}B^{2}\left( 1-\mathfrak{L}_{\mathfrak{F}}-%
\mathfrak{L}_{\mathfrak{FF}}B^{2}\right) k_{0}^{2}k_{\perp }^{2}\leqslant 0
\label{Poynting3}
\end{equation}%
for mode 3 (for the energy-momentum tensor components see Eqs. (61) and (58)
in (\cite{Shabus2011}), respectively,). It follows from DEC$_{\text{2 }}$ (%
\ref{Poynting2}) and WEC$_{\text{2}}$ (\ref{!+LGG}) that
\begin{equation}
\text{DEC}_{\text{2 }}\text{plus WEC}_{\text{2}}\text{: \ \ \ }\mathfrak{L}_{%
\mathfrak{GG}}\geqslant 0,  \label{LGG}
\end{equation}%
but we cannot establish the positivity of the two separate factors $%
\mathfrak{L}_{\mathfrak{FF}}B^{2}$ and $\left( 1-\mathfrak{L}_{\mathfrak{F}}-%
\mathfrak{L}_{\mathfrak{FF}}B^{2}\right) $ in (\ref{Poynting3}) appealing%
\footnote{%
We could, if we had appealed to the Maxwell dominance principle. But it is
not included into the Ellis-Hawking context.} to WEC and DEC.

On the contrary, by using our GVC causality conditions $0\leqslant \beta
_{2,3}\leqslant 1$ and PRC conditions we are able to reproduce all the above
inequalities WEC$_{3,\text{2}}$ (\ref{1-Lf}), (\ref{!+LGG}), and DEC$_{3,%
\text{2 }}$(\ref{Poynting2}), (\ref{Poynting3}) (as well as their
consequence (\ref{LGG}), of course) that follow from the energy conditions,
and, besides, obtain the two lacking ones.

As already said, WEC= PRC, hence Eqs. (\ref{1-Lf}), (\ref{!+LGG}) do not
need new derivation. In view of them the causality condition $\beta
_{3}\geqslant 0$ ( see Eqs. (\ref{betaG=0}) for $\beta ,$s) yields
\begin{equation}
\mathfrak{L}_{\mathfrak{FF}}\geqslant 0,  \label{LFF}
\end{equation}%
while the causality condition $\beta _{2}\geqslant 0$ reproduces (\ref{LGG}%
). Now, the causality condition $\beta _{3}\leqslant 1$ leads to the other
lacking inequality%
\begin{equation}
\left( 1-\mathfrak{L}_{\mathfrak{F}}-\mathfrak{L}_{\mathfrak{FF}%
}B^{2}\right) \geqslant 0,  \label{1-LFF}
\end{equation}%
while the causality condition $\beta _{2}\leqslant 1$ repeats already known
\ Eqs. (\ref{1-Lf}). It is notable that, the same as WEC=PRC, conditions $%
\beta _{3,2}\leqslant 1$ result in inequalities falling into the region of
Maxwell dominance, unlike the consequences of conditions $\beta
_{3,2}\geqslant 0,$ which are (\ref{LGG}) and (\ref{LFF}).

Sometimes also condition called null (NEC)
\begin{equation*}
N=t_{\mu \nu }n^{\mu }n^{\nu }\geqslant 0,
\end{equation*}%
where $n^{\nu }$ a future-pointing null-vector, is applied. We shall see
that it cannot supply the missing inequality (\ref{LFF}), either. Choosing $%
n_{\perp }=0$, and $n_{0}=n_{\parallel }=1$ and using expressions for
energy-momentum tensor from \cite{Shabus2011}) we find for mode 3:%
\begin{equation*}
N^{(3)}=\left( 1-\mathfrak{L}_{\mathfrak{F}}\right) (kn)^{2}+(kn)\mathfrak{L}%
_{\mathfrak{FF}}\left( nF^{2}k\right) \geqslant 0.
\end{equation*}%
Once the scalar product $\left( nF^{2}k\right) $ disappears, we are left
with the already established relation (\ref{1-LGG}). For mode 2, analogously,%
\begin{eqnarray}
N^{(2)} &=&\left( 1-\mathfrak{L}_{\mathfrak{F}}\right) (kn)^{2}+(kn)%
\mathfrak{L}_{\mathfrak{GG}}\left( n\widetilde{F}^{2}k\right) =  \notag \\
&=&\left( 1-\mathfrak{L}_{\mathfrak{F}}\right) (k_{0}-k_{\parallel
})^{2}+(k_{0}-k_{\parallel })\mathfrak{L}_{\mathfrak{GG}}B^{2}\geqslant 0.
\label{N2} \\
&&.  \notag
\end{eqnarray}%
On the dispersion curve the wave vector $k_{\mu }$ is space-like, and
time-directed in the farther-specialized frame where $k_{\parallel }=0$ (we
refer to explanation below Eq. (\ref{c+-})), that is $k_{0}>0.$ Hence $%
(k_{0}-k_{\parallel })>0$ in the special frame. Keeping this difference
small enough, we see that the previously established momentum-independent
inequality \ref{LGG} follows from (\ref{N2}). Therefore, Eq. (\ref{LFF})
remains unattainable by applying the null energy conditions, either.

We conclude with the remark that in order to make up the missing condition (%
\ref{LFF}), important for establishing the convexity of the Lagrangian, we
had to supplement DEC for mode 3 by either of the conditions $\beta
_{3}\geqslant 0$ or $\beta _{3}\leqslant 1$ taken from GVC. Recall in this
connection that mode 3 is the one that may be referred to as ordinary,
because its electric field is transverse to direction of propagation, $%
\left( \mathbf{e}_{3}\mathbf{k}\right) =0$ (also to external field, $\left(
\mathbf{e}_{3}\mathbf{B}\right) =0),$ whereas mode 2 is extraordinary in the
sense that its electric field has a nonzero longitudinal component $\left(
\mathbf{e}_{2}\mathbf{k}\right) =k_{\parallel }^{2}k^{2}=\beta
_{2}k_{\parallel }^{2}\mathbf{k}_{\perp }^{2}\geqslant 0,$ which is
nonnegative due to the corresponding GVC. (For descriptions of polarizations
of eigenmodes see \cite{shabad1975}$,$\cite{Trudy})$.$

\section{Acknowledgements}

The author is thankful to Profs. T. Adorno and D. Gitman, and Prof. B.
Voronov for fruitful discussions.

\end{document}